\documentclass[aps, reprint, prx, footinbib, superscriptaddress]{revtex4-2}

\usepackage[utf8]{inputenc}
\usepackage{float}
\usepackage{graphicx}  
\usepackage{physics}
\usepackage{esint}
\usepackage{amssymb}  
\usepackage{mathtools}
\usepackage{amsmath}
\usepackage{amsthm}
\usepackage{mathrsfs}
\usepackage{bm}
\usepackage{pifont}  
\usepackage{multirow}

\newcommand{\cmark}{\checkmark}
\newcommand{\xmark}{\ding{55}}

\usepackage[dvipsnames]{color,xcolor,colortbl}
\usepackage{hyperref}
\hypersetup{%
    colorlinks=true,
    linkcolor=RoyalPurple,
    citecolor=RoyalPurple,
    urlcolor=RoyalPurple,
    plainpages=false
    }
\usepackage[most]{tcolorbox}

\numberwithin{footnote}{section}
\numberwithin{equation}{section}

\begin{document}


\title{Incoherent Imaging with Spatially Structured Quantum Probes}

\author{Anthony J. Brady}\email{ajbrad4123@gmail.com}
\affiliation{Joint Center for Quantum Information and Computer Science, NIST/University of Maryland, College Park, MD, 20742, USA}
\affiliation{Joint Quantum Institute, NIST/University of Maryland, College Park, MD, 20742, USA}

\author{Zihao Gong}
\affiliation{Department of Electrical and Computer Engineering, University of Maryland, College Park, MD, 20742, USA}

\author{Alexey V. Gorshkov}
\affiliation{Joint Center for Quantum Information and Computer Science, NIST/University of Maryland, College Park, MD, 20742, USA}
\affiliation{Joint Quantum Institute, NIST/University of Maryland, College Park, MD, 20742, USA}

\author{Saikat Guha}
\affiliation{Department of Electrical and Computer Engineering, University of Maryland, College Park, MD, 20742, USA}

\begin{abstract}
    Incoherent imaging, including fluorescence and absorption microscopy, is often hindered by weak signals and resolution limits---notoriously, Rayleigh’s curse. We introduce a framework for incoherent quantum imaging in the perturbative (weak-signal) regime that leverages spatially structured quantum probes and multiplexed quantum readout to overcome these limitations. We propose a novel \emph{twin-beam echo} protocol that maps absorption and fluorescence onto distinct idler and signal modes that carry thermal-like signatures characteristic of passive imaging. This yields an explicit architecture for simultaneous absorption and fluorescence imaging at the quantum limit and allows passive-imaging methods to be directly repurposed for active quantum imaging. As a corollary, we show that subdiffraction absorption imaging with quantum probes evades Rayleigh's curse, whereas classical probes remain provably afflicted. Remarkably, the same protocol supports displacement-field reconstruction of multiple quadratures (e.g., oscillators' positions) and pertains to both conventional and subdiffraction imaging. We further identify Fock states in a structured spatial mode basis as a non-Gaussian alternative with comparable performance. Though primarily developed for optical imaging, our framework suggests broad applications to phononic and acoustic imaging, and to reconstructing forces, fields, or charge distributions with an array of mechanical oscillators.

\end{abstract}

\date{\today}

\maketitle


\section{Introduction}

The ability to spatially reconstruct incoherent signals---such as fluorescence from a complex emissive scene, the absorption profile of a material, or a stochastic field driving an array of sensors---is central to diverse fields and applications, ranging from biological and biomedical imaging~\cite{Taylor2016QuMetrologyBio,Defienne2024AdvQuImaging,Aslam2023nvBioMed} to condensed-matter physics~\cite{Casola2018nvCM,Rovny2024nvManyBody} and gravitational science~\cite{Schnabel2010:QuMetrologyLIGO,Jia2024:LigoSqz,Vermeulen2025:GQuEST}. However, conventional sensing and imaging often face two fundamental challenges: low measurement sensitivity in weak-signal regimes and (spatial) resolution limits. Quantum sensing~\cite{Degen2017:QuSensing, Pirandola2018:NatRvw} and imaging~\cite{Genovese2016:RvwQuImaging,Moreau2019NatRev,Albarelli2020Perspective} promise to overcome these limitations by exploiting quantum states and optimal measurements. For instance, seminal advances~\cite{Pirandola2018:NatRvw,Genovese2016:RvwQuImaging,Moreau2019NatRev,Albarelli2020Perspective} in quantum-optical sensing have shown that tailored quantum states, including twin beams, Fock states, and squeezed vacuum, can surpass classical shot-noise limits and significantly improve measurement sensitivity.

In similar fashion, quantum measurement techniques have revealed new strategies to bypass seemingly fundamental resolution barriers. The pioneering work of Tsang \emph{et al.}~\cite{Tsang2016Superresolution} showed that ``Rayleigh's curse,'' the apparent divergence in the estimation error of subdiffraction length scales, arises from suboptimal measurements strategies rather than a fundamental limit. This insight led to multiplexed quantum readout schemes, such as spatial-mode demultiplexing (SPADE)~\cite{Tsang2016Superresolution}, that achieve quantum-limited precision for subdiffraction imaging in passive (zero-probe-photon) settings~\cite{Tsang2016Superresolution, Tsang2019Starlight, Lupo2020PRL_QuLinearLimits, Wang2025:GaussMeasurements}. Similar resolution challenges and their quantum solutions extend beyond quantum optics, appearing in frequency-resolution limits of atomic and qubit sensors~\cite{Gefen2019NatComm_Resolution, Mouradian2021:ItermSignals, Dey2024EntangledFreqResol}.

\begin{figure}
    \centering
    \includegraphics[width=.8\linewidth]{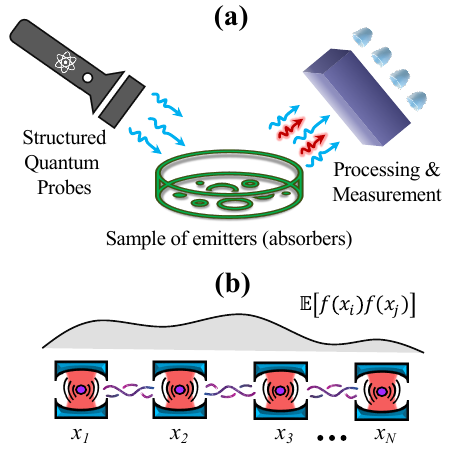}
    \caption{
    Quantum imaging with bosonic modes.
    (a) Structured quantum-optical probes interrogate a sample of emitters/absorbers. Spatial information about the sample is imprinted onto the probes through incoherent quantum processes (e.g., absorption and emission) and then extracted from the reflected light via quantum processing and measurements, such as spatial-mode demultiplexing (SPADE) and photon counting. 
    (b) Alternative setting subsumed by our framework where a stochastic field $f(x)$ drives an array of mechanical oscillators, such as trapped ions or nanomechanical resonators. A collective quantum state of the motional modes is prepared (squiggles represent entanglement) and measured to reveal properties hidden in the correlators $\mathbb{E}[f(x_i)f(x_j)]$. Collective heating (cooling) can likewise be estimated.
    }
    \label{fig:main}
\end{figure}

Yet many theoretical results in quantum imaging focus on the passive setting, where the imaging mode (e.g., the fluorescing Stokes channel) is initialized in vacuum, and often restrict to single-parameter estimation. Realistic applications, however, demand high-sensitivity detection of weak signals with fine spatial resolution in complex environments, where multiple parameters must be estimated simultaneously.
These challenges motivate us to explore quantum enhancements in multiparameter active imaging, where imaging modes are initialized in photon-rich quantum states. For instance, when interrogating a sample with non-vacuum probe light [Fig.~\ref{fig:main}(a)], both absorption and incoherent emission (a.k.a., fluorescence) processes generally occur together, encoding distinct physical information that must be jointly inferred. 

In a similar guise, reconstructing stochastic fields [Fig.~\ref{fig:main}(b)], such as fluctuating force fields driving trapped ions or mechanical resonators, amounts to displacement-field mapping across the sensor array, often via collective measurements and joint processing. By “displacement fields” we refer to spatially varying signals that displace the phase-space coordinates of the oscillators. These manifest as physical forces on motional degrees of freedom in mechanical settings and as quadrature shifts from background fields or other couplings in (microwave or optical) photonic settings. Though the connection between imaging and displacement sensing may not be immediately apparent, the problem setup and theoretical tools to address both turn out to be intimately connected. Developing strategies to address these multifaceted problems therefore presents outstanding opportunities in quantum imaging and sensing.

In this work, we introduce a \emph{universal quantum imaging module} that enhances both sensitivity and spatial resolution, whilst enabling simultaneous imaging of absorption and fluorescence signatures in the perturbative regime of weak signals.  As an additional modality, the same framework also applies to reconstruction of stochastic displacement fields. The central building block is a versatile technique based on pre- and post-processing via two-mode squeezing and its inversion, forming a twin-beam echo (a type of metrological Loschmidt echo~\cite{Macri2016:echoMetrology, Nolan2017:OptEchoReadout, Burd2019:SqzEcho,Colombo2022SATINecho}). The echo amplifies weak incoherent signals, separates fluorescence and absorption into distinct detection channels, and supports both coarse-grained and subdiffraction imaging. Key features of the emitter scene are extracted from photon-counting measurements in a structured spatial-mode basis. 

Remarkably, we find that our imaging module achieves quantum-optimal performance across a range of canonical imaging and estimation tasks. As a concrete example, we derive the ultimate limit for subdiffraction absorption imaging, i.e., estimating the separation between two closely spaced absorbers, and demonstrate that, in principle, a finite measurement sensitivity remains achievable as the separation tends to zero. Classical probes, on the other hand, intrinsically suffer from Rayleigh’s curse---that is, the measurement precision vanishes as the separation tends to zero---and thus require large photon numbers to maintain appreciable sensitivity in the subdiffraction regime. Our quantum imaging module overcomes this apparent limitation, achieving the ultimate measurement sensitivity and maintaining finite precision at zero separation with a fixed number of photons.

Previous single-mode studies have demonstrated quantum-enhanced estimation of either absorption (a.k.a., bosonic loss) or fluorescence (a.k.a., bosonic amplifier gain) individually using bright squeezed probes~\cite{Kamble2024:TBE} (see also Refs.~\cite{Monras2007LossEst,Aspachs2010:UnruhEst}). Our approach differs in that it supports simultaneous estimation of both components, extends naturally to multimode imaging, achieves quantum-enhanced resolution performance in the subdiffraction regime, and incorporates an additional modality for displacement-field reconstruction. Interestingly, our framework also connects to recent work linking incoherent optical imaging to stochastic waveform estimation, a variant of displacement sensing~\cite{Tsang2023NoiseSpectr}.

We further analyze perturbative loss, heating, and background noise and identify the noise channels to which a quantum advantage are first-order robust. For twin-beam echoes, robustness requires a sufficiently clean idler channel. We also assess the performance of structured Fock-state probes, which offer equivalent sensitivity enhancements to the twin-beam echoes, without ancillary modes (and thus without ancilla-assisted entanglement) but may be experimentally challenging to create.

The remainder of this paper is organized as follows. Section~\ref{sec:setup} introduces the theoretical imaging model. Section~\ref{sec:probes} presents the structured quantum probe protocols, where we introduce twin-beam echoes, analyze structured Fock-state probes, and discuss the shortcomings of single-mode squeezed-state probes. Section~\ref{sec:noise-effects} analyzes perturbative effects of noise and the robustness of our protocols. Section~\ref{sec:apps} surveys potential applications, and Section~\ref{sec:outlook} closes with an outlook. Several appendices provide supporting material, including: a microscopic motivation for our general imaging model (Appendix~\ref{app:model}); a brief review of passive imaging and Rayleigh's curse (Appendix~\ref{app:imaging}); detailed derivations of our twin-beam echo method (Appendix~\ref{app:echo}); and ultimate sensitivity limits via quantum Fisher information analyses of key imaging tasks (Appendix~\ref{app:channel-estimation}; see Table~\ref{tab:qfi_loss_gain_imaging} for a compact summary).

\begin{figure*}
    \centering
    \includegraphics[width=.85\linewidth]{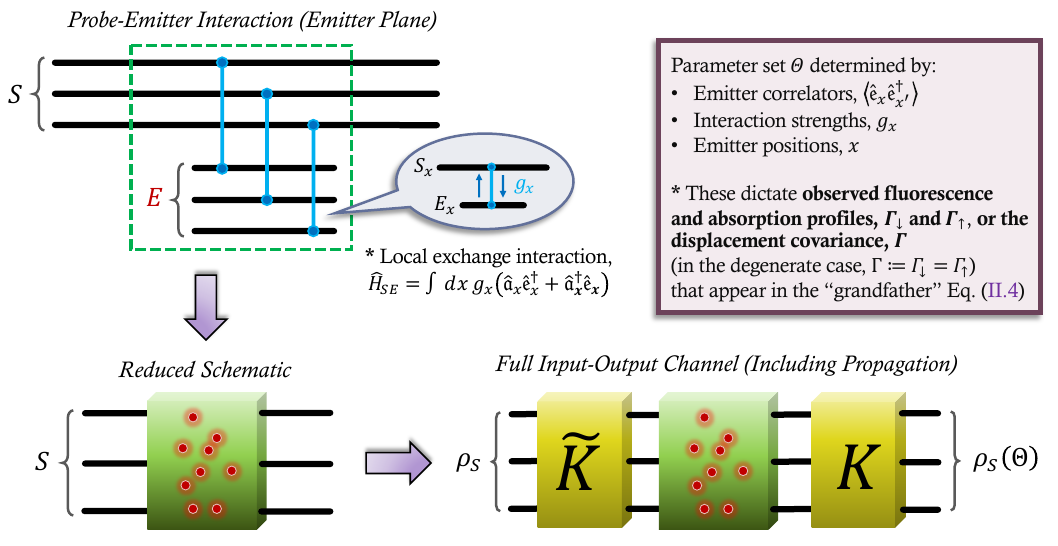}
    \caption{Pictorial construction of the quantum imaging model. 
    Local interactions between emitters (subsystems $E_x$) and probes (subsystems $S_x$) encode emitter properties onto the probes, such as absorption, fluorescence, and emitter positions.  
    Due to stochastic behavior of the emitters, this interaction induces incoherent evolution of the probes, described by a quantum-imaging channel acting on the probe state (reduced schematic).
    Propagation from the emitter plane to the detection plane (kernel $K$) spreads the encoded information across multiple ``pixel" modes on the collection plane, yielding the observed mutual coherence $\Gamma_{\downarrow}(u,u')$ and absorption profile $\Gamma_{\uparrow}(u,u')$ [or displacement-field covariance kernel $\Gamma(u,u')$]. Propagation from the probe source to the emitter plane (kernel $\widetilde{K}$) included for completeness.}
    \label{fig:interaction-schematic}
\end{figure*}

\section{Setup and Theoretical Model}
\label{sec:setup}

In this section, we describe the physical assumptions and the theoretical basis underlying our imaging model, focusing on scenarios that involve fluorescence and absorption. We provide microscopic derivations of the model in Appendix~\ref{app:model} for further motivation and support. See Fig.~\ref{fig:interaction-schematic} for an illustrative breakdown. Though phrased in quantum-optical imaging language, we also discuss how our formalism likewise encompasses displacement-field reconstruction under appropriate assumptions.

\paragraph*{\textbf{Assumptions and operating regime.}}
We consider a beam of probe photons interacting with a collection of absorbers/emitters confined to a single transverse plane. We model the probe as a nearly ideal paraxial pulse: a longitudinal plane wave of sufficiently long temporal duration, endowed with a structured transverse spatial profile.  Since the emitters lie on a single plane, only the transverse spatial-mode structure of the probe is relevant for imaging; for simplicity, we further restrict to a single transverse dimension (extension to two dimensions is straightforward). Hence, we describe the probe field on the emitter plane (with coordinate $x$) by the continuous set of localized field operators ${\hat{\mathfrak{a}}_x}$, and similarly, describe the field on the collection plane (coordinate $u$) by operators ${\hat{a}_u}$. These satisfy the bosonic commutation relations $\comm*{\hat{\mathfrak{a}}_x}{\hat{\mathfrak{a}}_{x'}^\dagger}=\delta(x-x')$ and $\comm*{\hat{a}_u}{\hat{a}_{u'}^\dagger}=\delta(u-u')$. We oft-refer to the coordinate degree of freedom as a ``pixel". See Appendix~\ref{app:model} for further details.

We suppose that the probe field does not significantly alter the emitter dynamics. Instead, the emitters are taken to be thermally populated or externally driven by (stochastic) background fields or environmental interactions. Consequently, the observed incoherent features of the emitters arise from this background activity, not from the probe itself, and the probe thus acts as a minimally invasive diagnostic of the emitters and their environment. For simplicity, we model the emitters as a Markovian bath (see Appendix~\ref{app:model}). Inclusion of non-Markovian effects or the use of ultra-short pulses (cf., Refs.~\cite{Kiilerich2019:InOutTheory, Albarelli2023:PulsedEst}) are possible but beyond the scope of this work.

We also assume relatively low-energy probes, such that the evolution of the probe field is perturbative. In this regime, single-photon processes governs the probe evolution. That is, the input-output quantum channel for the probe either subtracts (absorption) or adds (fluorescence) a single photon, with rates that can be stimulated by the number of photons in the probe. Multi-photon events are perturbatively suppressed. This operating regime is relevant in signal-starved conditions or weak-coupling scenarios (e.g., short interaction times), when probes must remain low energy to avoid disturbing the sample.

\begin{figure*}
    \centering
    \includegraphics[width=\linewidth]{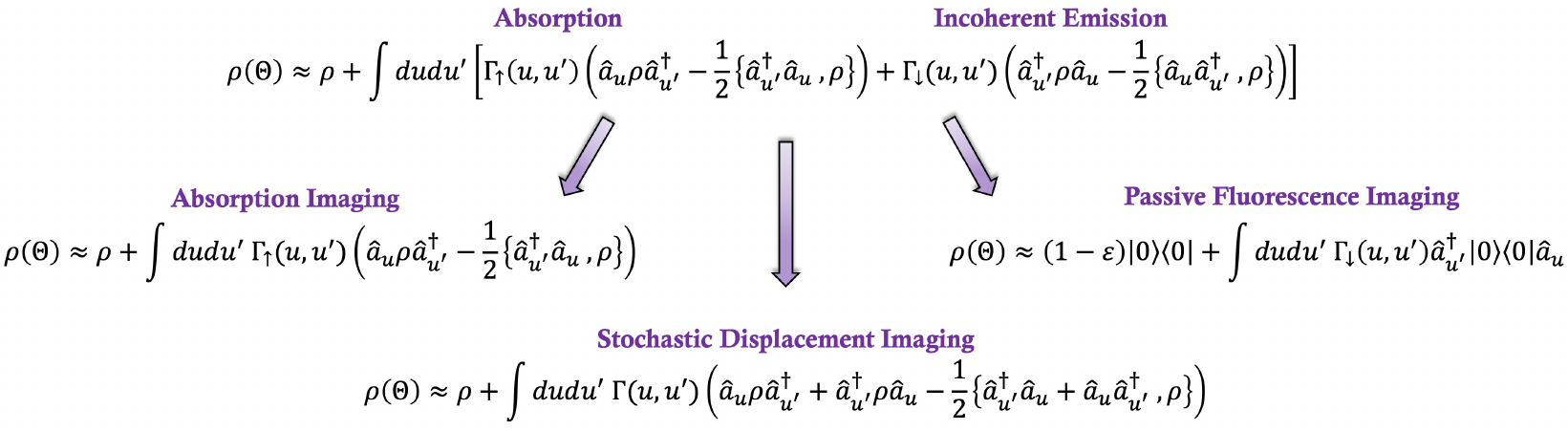}
    \caption{Lineage of the grandfather equation [Eq.~\eqref{eq:grandfather}]. Descendants include  absorption imaging, passive fluorescence imaging, and stochastic displacement imaging (a.k.a., displacement-field reconstruction).}
    \label{fig:grandfather}
\end{figure*}

\paragraph*{\textbf{Incoherent imaging model}}
As illustrated in Fig.~\ref{fig:interaction-schematic}, a structured quantum probe field propagates toward a scene of incoherent emitters or absorbers. Probe photons interact with the sample through two processes: Either probe photons are absorbed or incoherent excitations (fluorescence) are added to the probe field. In the microscopic model, these processes arise from a local exchange interaction that effectively swap quanta between the probe field and the emitters (see Fig.~\ref{fig:interaction-schematic}). The outgoing probe field carries the imprint of this interaction, which is subsequently processed by a quantum optical device and measured with a detector array. Measurement outcomes are then used to estimate a set of parameters, $\Theta=\{\theta_1,\theta_2,\dots\}$, such as the total brightness, absorption profile, separation distances between emitters/absorbers etc. The parameters are perturbatively encoded onto the probe, viz. $\rho\to \rho(\Theta)\coloneqq \rho +\delta\rho(\Theta)$, where $\rho$ is the input probe state and $\delta\rho(\Theta)$ is the perturbative piece that encodes $\Theta$. Our task is thus to model the weak interaction and derive an approximate form of the output probe state, $\rho(\Theta)$, which serves as the foundation of our sensing and imaging framework.

We access the emitters implicitly through the diagnostic probe and thus consider the evolution of the probe field alone following the weak interaction with the emitters. As elaborated in Appendix~\ref{app:model}, the post-interaction probe state, expressed on the emitter plane, is
\begin{align}
    \rho(\Theta) \approx \rho +\! \int\dd{x}&\dd{x'} \bigg[
        \gamma_\uparrow(x,x') \Big( \hat{\mathfrak{a}}_x \rho \hat{\mathfrak{a}}_{x'}^\dagger - \frac{1}{2} \acomm{\hat{\mathfrak{a}}_{x'}^\dagger \hat{\mathfrak{a}}_x}{\rho} \Big) \nonumber\\
        &+ \gamma_\downarrow(x,x') \Big( \hat{\mathfrak{a}}_{x'}^\dagger \rho \hat{\mathfrak{a}}_x - \frac{1}{2} \acomm{\hat{\mathfrak{a}}_x \hat{\mathfrak{a}}_{x'}^\dagger}{\rho} \Big)
    \bigg],
    \label{eq:rho_xplane}
\end{align}
where $\gamma_{\uparrow/\downarrow}(x,x')$ denote the absorption/emission correlators on the emitter plane, and $\hat{\mathfrak{a}}_x$ denotes the annihilation operator for probe field at position $x$ on the emitter plane. Our model permits spatial correlations across the emitter ensemble, e.g., collective emission or absorption effects, if present. For completely uncorrelated emitters, $\gamma_{\uparrow/\downarrow}(x,x')=\gamma_{\uparrow/\downarrow}(x)\delta(x-x')$, where $\gamma_{\uparrow/\downarrow}(x)$ denotes the local absorption/emission profile.

This formulation also clarifies the operational distinction between \emph{passive} and \emph{active} imaging adopted in this work: passive imaging uses a vacuum probe, $\rho=\dyad{0}$, whereas active imaging uses a non-vacuum probe state. These terms classify the probe state, rather than the physical origin of the detected signal. Consequently, thermal emission from an occupied emitter into otherwise unseeded modes (e.g., Stokes modes) constitutes passive imaging in our nomenclature, as in conventional fluorescence imaging.

The field modes $\hat{\mathfrak{a}}_x$ at the emitter plane are linearly related to the field modes $\hat{a}_u$ at the collection plane through the unitary propagation relation,
\begin{equation}
    \hat{\mathfrak{a}}_x = \int \dd{u} \, K(x,u) \hat{a}_u,
    \label{eq:bkw_prop}
\end{equation}
where $K(x,u)$ is the (backward) propagator. For simplicity, we assume ideal mode-matching between the probe-source and collection planes, so that the kernel $\widetilde{K}$ (see Fig.~\ref{fig:interaction-schematic}) that describes propagation from the probe-source to the emitter plane does not play an essential role here. In a more general treatment, both propagators, $\widetilde{K}$ and $K$, are necessary to relate the probe-source basis to the output collection-plane basis. Herein, we implicitly absorb $\widetilde{K}$ into the initial mode definition and thus replace it with identity assuming perfect mode-matching between the transmitter and collection device.

In many practical scenarios, we approximate $K(x,u)$ as a localized kernel $K(x,u) \approx \varphi(u-x)$, where $\varphi$ is the point spread function (PSF) that describes image blurring. A common approximation is a Gaussian profile,
\begin{equation}\label{eq:psf}
    \varphi(u) = \frac{e^{-u^2/4\sigma^2}}{(2\pi \sigma^2)^{1/4}}.
\end{equation}
In optical microscopy, one has $\sigma \approx \lambda f/(\pi w)$, with wavelength $\lambda$, focal length $f$, and aperture radius $w$~\cite{Zhou2019zLocalization}. This approximation captures the key constraint in subdiffraction imaging, which is the finite resolution set by $\sigma$.

Substituting the propagation relation~\eqref{eq:bkw_prop} into Eq.~\eqref{eq:rho_xplane} yields the probe state at the collection plane,
\begin{align}
    \rho(\Theta) \approx \rho + \int\dd{u}\dd{u'} \bigg[
        \Gamma_\uparrow(u,u') \Big( \hat{a}_u \rho \hat{a}_{u'}^\dagger - \frac{1}{2} \acomm{\hat{a}_{u'}^\dagger \hat{a}_u}{\rho} \Big) \nonumber\\
        + \Gamma_\downarrow(u,u') \Big( \hat{a}_{u'}^\dagger \rho \hat{a}_u - \frac{1}{2} \acomm{\hat{a}_u \hat{a}_{u'}^\dagger}{\rho} \Big)
    \bigg],
    \label{eq:grandfather}
\end{align}
where
\begin{align}
    \Gamma_\uparrow(u,u') &= \int\dd{x}\dd{x'} \, \gamma_\uparrow(x,x') K(x, u)K^*(x', u'), \label{eq:abs-kernel}\\
    \Gamma_\downarrow(u,u') &= \int\dd{x}\dd{x'} \, \gamma_\downarrow(x,x') K(x, u)K^*(x', u')\label{eq:mutual-coh}
\end{align}
define the absorption kernel and mutual coherence, $\Gamma_{\uparrow}$ and $\Gamma_{\downarrow}$, of the output field~\footnote{We only consider emission into modes matched to the transmitter and collection device; fluorescence into other modes is inevitably lost.}. We refer to Eq.~\eqref{eq:grandfather} as the ``grandfather'' equation, as its lineage encompasses many well-studied imaging and sensing models in suitable limits (see Fig.~\ref{fig:grandfather}).

For negligible emission ($\Gamma_\downarrow\approx 0$), the model reduces to an absorption-only channel, relevant for transmission and absorption imaging~\cite{Brambilla2008LossImaging,Brida2010SubShotImaging,Samantaray2017QuWideMicroscope,Moreau2017:AbsoluteAbsorp} (and directly related to bosonic loss estimation~\cite{Monras2007LossEst,Adesso2009LossEst}). When the input state is vacuum ($\rho=\dyad{0}$), only the emission channel contributes, thereby recovering well-studied passive imaging models in the single-photon regime (see, e.g., Refs.~\cite{Tsang2016Superresolution,Lupo2016PRL_SubwaveImaging,Lupo2020PRL_QuLinearLimits}). When emission and absorption rates are balanced ($\Gamma_\downarrow\approx \Gamma_\uparrow$), as in the classical high-temperature limit of a thermal bath or under engineered symmetric noise, the evolution formally reduces to a multimode stochastic displacement channel, with zero mean displacement and covariance kernel $\Gamma(u,u')$. This channel is broadly pertinent to, for instance, force-field reconstruction [see Fig.~\ref{fig:main}(b)] with optomechanical systems~\cite{Xia2023OmechDQS, Brady2023OmechArray}, imaging random charge or electric-field distributions with an ensemble of trapped ions~\cite{Brownnutt2015:IonEnoise, Gilmore2021IonEFieldQSN}, sensing background microwave fields with a network of cavities~\cite{Brady2022:DMsearch}, and more generally, distributed sensing of quadrature field fluctuations~\cite{Brady2024CorrNoiseQSN}. In contrast to the propagating-light setting, where probes travel to the emitters and back, here one prepares a many-body state of the oscillators (e.g., ions confined in trap), allows the force or field to act on the system, and then performs a collective measurement. We point to Section~\ref{sec:apps} for a myriad of potential applications and further discussion. 

We remark that the perturbative treatment of the effective probe evolution [Eq.~\eqref{eq:grandfather}] holds provided that
\begin{align}
    \int\dd{x}\dd{x'}\gamma_{\downarrow}(x,x')\Tr\{\rho\hat{\mathfrak{a}}_x\hat{\mathfrak{a}}_{x'}^\dagger\} &\ll 1, \\
    \int\dd{x}\dd{x'}\gamma_{\uparrow}(x,x')\Tr\{\rho\hat{\mathfrak{a}}_{x'}^\dagger\hat{\mathfrak{a}}_x\} &\ll 1,
\end{align}
which ensures that single-jump events dominate while multi-jump events are parametrically suppressed. Physically, this corresponds to signal-starved settings and minimally invasive probes. For metrological tasks, at sufficiently high statistical precision, higher-order corrections must be retained to prevent perturbative model bias from exceeding the statistical estimation error.

\paragraph*{\textbf{Structured mode expansion.}}

For various imaging tasks, it is often useful (and necessary) to leverage a structured mode basis $\Psi = \{\psi_k\}$, such as the Hermite-Gaussian (HG) modes. Consider the expansion of the mutual coherence in this basis,
\begin{equation}
    \Gamma_{\downarrow}(u,u') = \sum_{\ell,k} \Gamma_{\downarrow,\ell k}^{[\Psi]} \psi_\ell^*(u') \psi_k(u),
\end{equation}
with similar expressions for the absorption kernel, $\Gamma_\uparrow$, and define the mode operators $\hat{\psi}_k = \int \dd{u} \psi_k(u) \hat{a}_u$. From which we rewrite Eq.~\eqref{eq:grandfather} as
\begin{align}
    \rho(\Theta) \approx \rho + \sum_{\ell,k} \bigg[
        \Gamma_{\uparrow,\ell k}^{[\Psi]} \Big( \hat{\psi}_k \rho \hat{\psi}_\ell^\dagger - \frac{1}{2} \acomm{\hat{\psi}_\ell^\dagger \hat{\psi}_k}{\rho} \Big) \nonumber\\
        + \Gamma_{\downarrow,\ell k}^{[\Psi]} \Big( \hat{\psi}_\ell^\dagger \rho \hat{\psi}_k - \frac{1}{2} \acomm{\hat{\psi}_k \hat{\psi}_\ell^\dagger}{\rho} \Big)
    \bigg].
    \label{eq:structured-grand}
\end{align}
This equation serves as the foundation for quantum enhanced subdiffraction imaging with structured probes. The choice of mode basis is crucial for imaging performance, as emphasized by Tsang \emph{et al.}~\cite{Tsang2016Superresolution}, and the optimal basis generally depends on the specific problem and the PSF.

\section{Imaging with Structured Quantum Probes}
\label{sec:probes}

\begin{figure}[t]
    \centering
    \includegraphics[width=\linewidth]{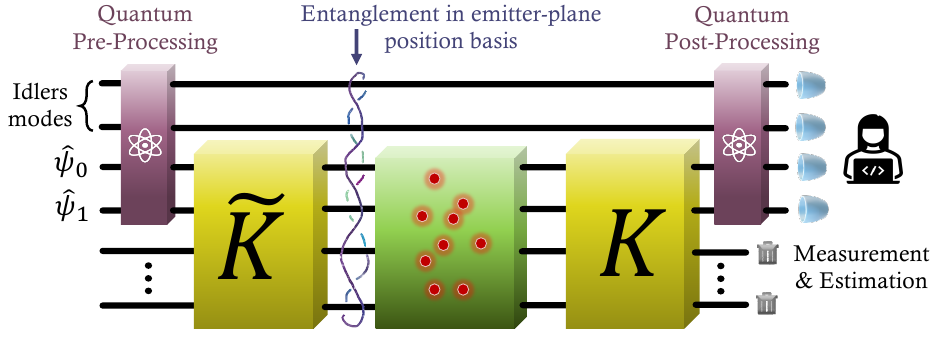}
    \caption{Quantum circuit for structured imaging. Quantum probes (twin beams or Fock states) are prepared in a structured mode basis $\Psi=\{\psi_k\}$. Idler modes (if present) are retained in storage, while the signal modes propagate to the emitter plane (via propagator $\widetilde{K}$), interact with the sample, then propagate to the collection plane (via propagator $K$) for processing and measurement. Quantum processing refers to energy non-conserving operations, such as squeezing, as well as linear-optical operations, such as SPADE.}
    \label{fig:probe-schematic}
\end{figure}

In this section, we introduce the universal quantum imaging module for simultaneous absorption and fluorescence imaging in the perturbative regime of weak signals. We provide a high-level circuit schematic in Fig.~\ref{fig:probe-schematic}, in which structured quantum probes are prepared in a structured mode basis, propagate toward the scene, undergo interaction with the sample, and are subsequently processed and measured at a collection device to perform estimation and imaging tasks. We focus primarily on two classes of probes: entangled twin-beam states and structured Fock-state probes. Both can be prepared in a spatial mode basis adapted to the imaging task and combined with mode-selective detection strategies, such as spatial-mode demultiplexing (SPADE) and photon counting, to achieve quantum enhancements in multiparameter estimation. We also discuss the limitations of single-mode squeezed states for imaging and assess the impact of Gaussian noise.

We summarize optimal sensitivity bounds in Table~\ref{tab:qfi_loss_gain_imaging} (and derived in Appendix~\ref{app:channel-estimation}). These provide strong evidence that the proposed imaging strategies operate at or near the ultimate quantum limits, substantially outperforming classical approaches in relevant regimes.

\subsection{Twin-beam echoes}
\label{sec:twin-beam-imaging}

\begin{figure}
    \centering
    \includegraphics[width=.9\linewidth]{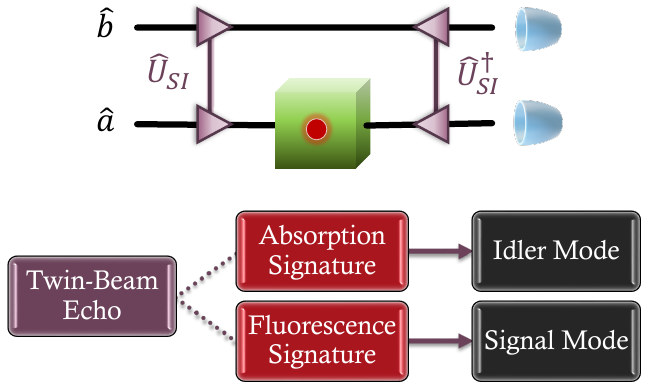}
    \caption{Twin-beam echo. A metrological Loschmidt echo~\cite{Macri2016:echoMetrology} via two-mode squeezing between signal ($S$, annihilation operator $\hat a$) and idler ($I$, annihilation operator $\hat b$) maps absorption signatures to the idler while maintaining fluorescence signatures on the signal.}
    \label{fig:toyecho}
\end{figure}

\begin{figure}
    \centering
    \includegraphics[width=\linewidth]{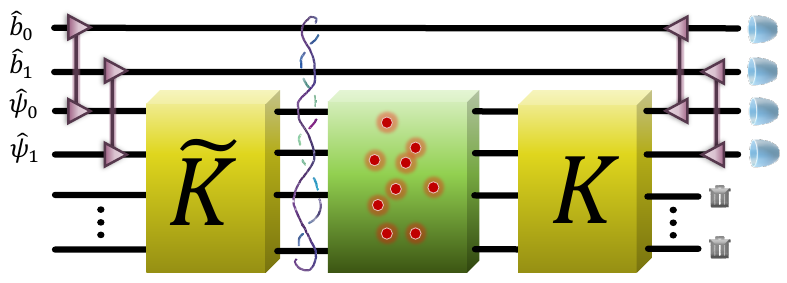}
    \caption{Structured twin-beam echoes. Independent echo sequences tailored to the structured mode basis ${\Psi=\{\psi_k\}}$. Two-mode squeezing operations (paired triangles) entangle signal-idler pairs. The idler modes ($\hat{b}_k$) are held in storage while signal modes ($\hat{\psi}_k$) propagate to the emitter plane and interact with the emitters. After interaction, the signal modes propagate to the collection device, and the initial two-mode squeezing is reversed (inverted paired triangles).}
    \label{fig:echo}
\end{figure}

One of the central results of this work is an imaging protocol based on \emph{twin-beam echoes}, in which pairs of initially entangled signal-idler modes, created via two-mode squeezing, are nearly disentangled by the inverse process after the probes interact with the emitter scene. This sequence realizes a metrological Loschmidt echo~\cite{Macri2016:echoMetrology, Nolan2017:OptEchoReadout, Colombo2022SATINecho} and provides a powerful means to selectively isolate and amplify weak incoherent signatures. We depict a simplified single-pixel schematic in Fig.~\ref{fig:toyecho}, with the full multimode configuration in Fig.~\ref{fig:echo}. 

We first present a toy example of simultaneous loss and gain estimation to build intuition for the basic echo mechanism before generalizing to the full imaging scenario. As another demonstration, we show how the twin-beam echo facilitates quantum-enhanced subdiffraction imaging, achieving resolution beyond the Rayleigh limit with improved sensitivity, and more broadly supports simultaneous absorption and fluorescence imaging. The twin-beam echo thus serves as a key building block of our quantum imaging module, providing a versatile and near-optimal tool for both conventional and subdiffraction imaging.

\paragraph*{\textbf{Toy echo example.}}
As a warm-up, we consider a ``zero-dimensional'' imaging problem, where absorption and fluorescence are localized to a single pixel and fully described by two parameters, $\Gamma_{\uparrow/\downarrow}$. We aim to estimate these parameters simultaneously.

Let $\hat{a}$ denote the lone signal mode and $\hat{b}$ the idler. See Fig.~\ref{fig:toyecho}. Suppose we prepare a twin beam, $\rho_{SI}=\hat{U}_{SI}\dyad{0}_{SI}\hat{U}_{SI}^\dagger$, where $\hat{U}_{SI}$ generates two-mode squeezing between the signal and idler. Only the signal mode interacts with the emitter, while the idler is maintained in storage (e.g., propagates along a lossless path or stored in quantum memory if necessary). From Eq.~\eqref{eq:grandfather}, the output after emitter-probe interaction is simply
\begin{align}
    \rho_{SI}(\Theta)\approx \rho_{SI} &+ \Gamma_{\uparrow}\Big(\hat{a}\rho\hat{a}^\dagger -\frac{1}{2}\acomm*{\hat{a}^\dagger\hat{a}}{\rho}\Big) \nonumber\\
    &+ \Gamma_{\downarrow}\Big(\hat{a}^\dagger\rho\hat{a}-\frac{1}{2}\acomm*{\hat{a}\hat{a}^\dagger}{\rho}\Big).
\end{align}
We then apply the reversed operation $\hat{U}_{SI}^\dagger$, producing the final state $\rho_{SI}^{\text{echo}} = \hat{U}_{SI}^\dagger \rho_{SI}(\Theta) \hat{U}_{SI}$. In the weak-coupling regime, this operation reverses the squeezing on the vacuum component while mapping fluorescence (incoherent excitations of the signal mode) and absorption (loss) into distinguishable excitations on the signal and idler modes, respectively. 

In the Heisenberg picture, the echo enacts the operator transformation $\hat{U}_{SI}^\dagger \hat{a} \hat{U}_{SI} = \cosh r\,\hat{a}+\sinh r\,\hat{b}^\dagger$, resulting in the echoed state
\begin{align}
    \rho_{SI}^{\text{echo}} \approx &(1 - \Gamma_I - \Gamma_S)\dyad{0}_S\otimes\dyad{0}_I \nonumber \\
    &+ \Gamma_I \dyad{0}_S\otimes\dyad{1}_{I} 
    + \Gamma_S \dyad{1}_{S}\otimes\dyad{0}_I \nonumber \\ 
    &+ C\dyad{1}{0}_S\otimes\dyad{1}{0}_I+C^*\dyad{0}{1}_S\otimes\dyad{0}{1}_I,
    \label{eq:toy-echo-state}
\end{align}
where $\ket{1}_S=\hat{a}^\dagger\ket{0}_S$ and $\ket{1}_I=\hat{b}^\dagger\ket{0}_I$ are single-photon states on the signal and idler modes, $\Gamma_S = \Gamma_\downarrow \cosh^2 r$, and $\Gamma_I = \Gamma_\uparrow \sinh^2 r$. Cross-terms (involving the unspecified coefficient $C$) encode correlations between the signal and idler but are unimportant for our purposes.

Crucially, the echo separates fluorescence and absorption into distinguishable excitations on signal and idler channels; see the flowchart in Fig.~\ref{fig:toyecho}. To see this more directly, suppose we restrict to separable measurements on signal and idler sectors. Then, the measurement-averaged state factorizes as $\bar{\rho}_{SI}^{\,\rm echo}\approx\rho_S^{\rm echo}\otimes\rho_I^{\rm echo}$, with $\rho_{S}^{\rm echo}=(1-\Gamma_S)\dyad{0}_S+\Gamma_S\dyad{1}_S$ and ${\rho_{I}^{\rm echo}=(1-\Gamma_I)\dyad{0}_I+\Gamma_I\dyad{1}_I}$. The amplified single-photon probabilities $p_S=\Gamma_S$ and $p_I=\Gamma_I$ directly encode the fluorescence and absorption rates. 

Due to the separability of the (measurement-averaged) output state, we can \emph{simultaneously} estimate $\Gamma_\uparrow$ and $\Gamma_\downarrow$ by independently counting photons on the signal and the idler. Moreover, this strategy saturates the ultimate single-parameter quantum limits for each; see Table~\ref{tab:qfi_loss_gain_imaging} for a quick summary and Appendix~\ref{app:sensitivity-limits} for further details. Using the standard Fisher-information formula for binary click/no-click outcomes (with probabilities $p_0= 1-p_1$ and $p_1\approx\Gamma_{S/I}\ll 1$), we find the classical Fisher informations of the imaging protocol,  
\begin{align}
    F_C(\Gamma_\downarrow)&\approx\frac{\cosh^2r}{\Gamma_\downarrow},\\
    F_C(\Gamma_\uparrow)&\approx\frac{\sinh^2r}{\Gamma_\uparrow},
\end{align}
which achieves quantum-optimal estimation of absorption (a.k.a., bosonic loss~\cite{Monras2007LossEst,Adesso2009LossEst,Nair2018:LossEst,Gong2023:TransmitSens}) and incoherent emission (a.k.a., bosonic amplification~\cite{Aspachs2010:UnruhEst}) in the perturbative regime; see Table~\ref{tab:qfi_loss_gain_imaging}. In other words, for a fixed (average) number of photons, we can simultaneously estimate $\Gamma_\downarrow$ and $\Gamma_\uparrow$ at the ultimate quantum limit. The twin-beam echo thus generalizes single-parameter results to joint estimation without sacrificing optimality (see also~Ref.~\cite{Nair2023:CovChSensing} for related multiparameter bounds) and, importantly, provides an explicit measurement strategy.

A recent study by Kamble \emph{et al.}~\cite{Kamble2024:TBE} considered a closely related echo protocol for estimating either absorption \emph{or} gain in a single-mode optical channel using bright two-mode squeezed states (see also Ref.~\cite{Zhang2025:tbeWeakAbs} in the weak-absorption limit)~\footnote{Note that we can view the twin-beam echo as a special case of an $SU(1,1)$ interferometer~\cite{Marino2012:SU11Loss,Kamble2024:TBE}, where the first and second squeezing operations play the role of active beam-splitters in a time-reversed configuration.}. On the other hand, our framework (\emph{i}) enables \emph{simultaneous} estimation of absorption and fluorescence via separation into distinct detection channels and (\emph{ii}) generalizes naturally to multimode imaging as we show below. The novelty therefore lies in these joint-estimation and structured-multimode extensions, rather than in the underlying echo sequence itself. Furthermore, we also analyze noise effects, such as loss and parasitic heating (see Section~\ref{sec:noise-effects}).

\paragraph*{\textbf{Structured twin-beam echoes.}}
We now extend the single-pixel echo to the full multimode configuration, allowing us to address complex emitter scenes and various imaging tasks. See Fig.~\ref{fig:echo} for a depiction. This extension is surprisingly straightforward but algebraically tedious, so we simply state the main result here and defer a full proof (using the Gaussian covariance formalism) to Appendix~\ref{app:echo}.

Suppose that we implement separable twin-beam echoes between signal modes in the structured basis $\Psi = \{\psi_k\}$ and an arbitrary set of idler modes $\{\hat{b}_k\}$. Let $r_k$ denote the independent squeezing strengths for each twin beam (see Fig.~\ref{fig:echo}). Furthermore, we restrict to separable measurements on signal and idler sectors. Under these conditions, the (measurement-averaged) echoed states become 
\begin{align}
    \rho_S^{\text{echo}} &\approx (1 - \Gamma_S)  \dyad{0}_S  \nonumber \\
    &\quad + \sum_{k,\ell} \Gamma_{\downarrow, k\ell}^{[\Psi]} \cosh r_k \cosh r_\ell \,  \hat{\psi}_\ell^\dagger  \dyad{0}_S  \hat{\psi}_k, \label{eq:signal-echo}\\
    \rho_I^{\text{echo}} &\approx (1 - \Gamma_I)  \dyad{0}_I \nonumber\\
    &\quad + \sum_{k,\ell} \Gamma_{\uparrow, k\ell}^{[\Psi]} \sinh r_k \sinh r_\ell  \hat{b}_\ell^\dagger  \dyad{0}_I  \hat{b}_k,\label{eq:idler-echo}
\end{align}
with amplified rates $\Gamma_S =\sum_k \Gamma_{\downarrow, kk}^{[\Psi]} \cosh^2 r_k $ and $\Gamma_I =\sum_k \Gamma_{\uparrow, kk}^{[\Psi]} \sinh^2 r_k$. Notably, the idler modes are arbitrary. The only requirement is that they be entangled with appropriate signal modes in the basis $\Psi$.

The resulting echoed states are independent incoherent single-photon states (i.e., approximate thermal states in the single-photon approximation) that exhibit intra-sector correlations, analogous to the quantum model of passive fluorescence imaging (see Fig.~\ref{fig:grandfather}) but with boosted single-photon detection probabilities due to the squeezing factors. Consequently, the quantum theory and techniques developed for passive imaging carry over directly to this active-probe setting.

\paragraph*{\textbf{Subdiffraction imaging.}}
We now consider
the canonical problem in subdiffraction imaging of estimating the separation $d$ between two uncorrelated, identical point-source emitters in the subdiffraction regime ($d \ll \sigma$)~\cite{Tsang2019Starlight}. This simple problem serves as a testbed for structured twin-beam echoes, illustrating how the protocol amplifies single-photon detection probabilities---thus improving overall signal-to-noise performance---while simultaneously achieving sub-Rayleigh spatial resolution. For an overview of subdiffraction imaging, see Appendix~\ref{app:imaging}.

We first focus on fluorescence signatures and thus restrict our analysis to the echoed signal modes $\rho_S^{\rm echo}$ [Eq.~\eqref{eq:signal-echo}]. We then analyze absorption signatures that reside on the echoed idlers [Eq.~\eqref{eq:idler-echo}].

Consider two uncorrelated emitters of equal brightness separated by a distance $d \ll \sigma$ (subdiffraction regime); physically, we mean that the emitter separation is smaller than the characteristic scale of the imaging system, yet larger than any correlation length of the emitter system. We also assume the centroid is known with  $x_1+x_2=0$. The mutual coherence at the collection plane is then (see Appendix~\ref{app:imaging} for details):
\begin{equation}
    \Gamma_{\downarrow}(u,u') = \frac{\varepsilon}{2} \sum_{m=1}^2 \varphi(u-x_m) \varphi(u'-x_m),
\end{equation}
where $\varepsilon$ is the total (observed) brightness and $\varphi(u-x)$ is the Gaussian PSF~\eqref{eq:psf}. In this case, the optimal mode basis is the HG basis $\Psi = \{\psi_0, \psi_1, \ldots\}$, where only the first two elements (the fundamental and antisymmetric modes, $\psi_0$ and $\psi_1$) contribute significantly. To leading order in $d/\sigma$, one may readily show that the coefficients of the mutual coherence in the HG basis are (see Appendix~\ref{app:imaging} for details)
\begin{align}
    \Gamma_{\downarrow,00}^{[\Psi]} &= \varepsilon \left( 1 - \frac{d^2}{16\sigma^2} \right),\\
    \Gamma_{\downarrow,11}^{[\Psi]} &= \frac{\varepsilon d^2}{16\sigma^2}, \\
    \Gamma_{\downarrow,01}^{[\Psi]} &= \Gamma_{\downarrow,10}^{[\Psi]} = 0,
\end{align}
with higher-order terms of negligible importance. For brevity, we take equal squeezing strengths, $r_0=r_1=r$, for the twin-beams associated with $\psi_0$ and $\psi_1$. This corresponds to a total mean signal photon number $N_S=2\sinh^2(r)$ (and total signal-plus-idler photon number $2N_S$). Defining the amplified brightness $\tilde{\varepsilon}=\varepsilon \cosh^2(r)$, the echoed signal state [Eq.~\eqref{eq:signal-echo}] simplifies to
\begin{multline}
    \rho_S^{\rm echo} \approx (1 - \tilde{\varepsilon})\dyad{0} + \tilde{\varepsilon} \left(1 - \frac{d^2}{16\sigma^2} \right) \hat{\psi}_0^\dagger \dyad{0} \hat{\psi}_0 
     \\
    + \frac{\tilde{\varepsilon} d^2}{16\sigma^2} \hat{\psi}_1^\dagger \dyad{0} \hat{\psi}_1.
\end{multline}
The twin-beam echo thus amplifies the photon-count probabilities in each mode by $\cosh^2 r$, simultaneously improving the signal-to-noise in brightness ($\varepsilon$) and separation ($d$) estimation. In point of fact, we calculate the classical Fisher informations to be
\begin{align}
    F_C(\varepsilon)&\approx \frac{\cosh^2r}{\varepsilon} \label{eq:epsilon-FC}\\ 
    F_C(d/\sigma)&\approx \frac{\varepsilon \cosh^2r}{4}.
\end{align}
The Fisher information for $d$ (in units of $\sigma$) is $\cosh^2r$ times that of the passive (zero probe photons) quantum Fisher information~\cite{Tsang2016Superresolution,Lupo2016PRL_SubwaveImaging}; see also Table~\ref{tab:qfi_loss_gain_imaging}. Moreover, the parameters $\varepsilon$ and $d$ can be estimated simultaneously: $\varepsilon$ can be estimated from the total photon counts on $\psi_0$ and $\psi_1$ and $d$ can be inferred from $\psi_1$ alone.

Likewise, absorption features can be reconstructed from measurements on the idler modes. Let $\gamma_{\uparrow}$ denote the (normalized) absorption rate per absorber. Measurements on the idler sector imply
\begin{align}
    F_C(\gamma_{\uparrow})&\approx \frac{2\sinh^2r}{\gamma_{\uparrow}}, \label{eq:abs-FC}\\ 
    F_C(d/\sigma)&\approx \frac{\gamma_{\uparrow}}{2} \sinh^2r.
\end{align}
Astonishingly, subdiffraction absorption imaging is free from Rayleigh's curse, similar to passive fluorescence imaging~\cite{Tsang2016Superresolution}. The crucial distinction in absorption imaging is that a probe is necessary---i.e., photons must be initially present to be absorbed. While classical probes (convex combinations of coherent states~\cite{Glauber1963, Sudarshan1963}) are inherently afflicted by Rayleigh (Appendix~\ref{app:classical-limits}), we demonstrate that quantum probes avert the curse entirely. See Table~\ref{tab:qfi_loss_gain_imaging} for quick reference and Appendix~\ref{app:subdiff-qfi} for details.

Altogether, we have shown that fluorescence and absorption scenes can be simultaneously imaged at subdiffraction scales with quantum-enhanced sensitivity, thereby improving overall signal-to-noise performance whilst circumventing Rayleigh's resolution limit. There is no measurement incompatibility at this order, since the full parameter set $(\varepsilon,\gamma_\uparrow,d)$ can be estimated jointly from commuting, mode-resolved photon-count statistics on the signal and idler sectors. The relevant trade-off is instead the allocation of a fixed photon budget $N_S$ among spatial modes to estimate these parameters most efficiently. The same considerations apply to structured Fock-state probes.

Although we presently focus on the canonical two-point-source problem, we expect that the analysis is extendable to more complex scenarios by adapting existing passive-imaging methods to the current framework. We briefly discuss many related problems in Appendix~\ref{app:beyond-problems}.

\paragraph*{\textbf{Conventional imaging.}}

The twin-beam echo also enhances signal-to-noise in conventional direct-detection imaging, where resolving subdiffraction features may not be required. In this setting, the task is to reconstruct a coarse-grained image of the scene by directly measuring the intensity at each pixel $u$ on the collection plane. The protocol proceeds as follows. 

Suppose the two-mode squeezing operations in the echo are implemented between spatially conjugate points of the signal and idler fields, $\hat{a}_u$ (signal) and $\hat{b}_{-u}$ (idler), with fixed squeezing strength $r_u$ per $u$. We assume that the transmitter plane (containing the initial two-mode squeezer) is mode-matched to the collection device (containing the inverted two-mode squeezer), allowing us to describe the protocol using a single spatial coordinate $u$; see Refs.~\cite{Brambilla2008LossImaging,Brida2010SubShotImaging,Samantaray2017QuWideMicroscope,Moreau2017:AbsoluteAbsorp} for physical realization of point-wise entangled twin-beams via Spontaneous Parametric Down Conversion (SPDC). After the echo, direct detection (via pixelated photodetectors) of the signal and idler modes yields per-pixel intensities
\begin{align}
    \expval{\hat{a}_u^\dagger\hat{a}_u} &\propto \cosh^2 r_u  \Gamma_{\downarrow}(u,u), \\
    \expval*{\hat{b}_{u}^\dagger\hat{b}_{u}} &\propto \sinh^2 r_{-u}  \Gamma_{\uparrow}(-u,-u).
\end{align}
Thus, the echo protocol yields amplified direct-detection signatures of both fluorescence and absorption profiles, which appear separately on the signal and idler modes. Importantly, this scheme remains compatible with conventional intensity-imaging hardware while offering improved sensitivity through engineered probe states---extending SPDC-based absorption imaging~\cite{Brambilla2008LossImaging,Brida2010SubShotImaging,Samantaray2017QuWideMicroscope,Moreau2017:AbsoluteAbsorp} to a unified fluorescence and absorption imaging module.

\paragraph*{\textbf{Reconstructing displacement fields.}} 
We analyze stochastic displacement-field reconstruction using the same twin-beam echo strategy introduced for fluorescence and absorption imaging. This resonates with recent observations by Tsang~\cite{Tsang2023NoiseSpectr}, who drew a formal analogy between incoherent optical imaging and stochastic waveform estimation (i.e., spectroscopy of mechanical displacement fluctuations~\cite{Ng2016PRA_NoiseSpectro}).  

Consider the special case where absorption and fluorescence are balanced, $\Gamma_\uparrow = \Gamma_\downarrow \eqqcolon \Gamma$. In this limit, the grandfather equation reduces to a stochastic multimode quadrature-displacement channel, where quadrature displacements occur on the emitter plane and are then detected on the collection plane. As seen on the collection plane, displacements are described by the unitary operator $\hat{D}(\vec{\beta}) = \exp\!\big(\int \dd{u}\,\beta_u \hat{a}_u^\dagger - \beta_u^*\hat{a}_u\big)$, with $\beta_u\sim\mathcal{CN}(0,\Gamma(u,v))$ denoting the complex-Gaussian random field with covariance kernel $\mathbb{E}[\beta_u \beta_v^*] = \Gamma(u,v)$. Expressed in quadrature variables, $\beta_u = (q_u + i p_u)/\sqrt{2}$, we have $\mathbb{E}[q_u q_v] = \mathbb{E}[p_u p_v] = \Gamma(u,v)$, with all other correlators vanishing. [The analysis generalizes straightforwardly to asymmetric quadrature variances, e.g., purely positional displacements.] Physically, the displacement model corresponds to an incoherent process that stochastically shifts the oscillators' phase-space coordinates. Examples include mechanical fluctuations in optomechanical resonators due to background (semi-classical) forces~\cite{Xia2023OmechDQS}, fluctuating electric fields causing positional jitter in trapped ions~\cite{Brownnutt2015:IonEnoise, Gilmore2021IonEFieldQSN}, or random microwave drives in superconducting cavities~\cite{Agrawal2024:FockDM} (see Section~\ref{sec:apps} for more discussion).

In this setting, we estimate properties of the displacement covariance $\Gamma$ using photon counts from both signal and idler modes after the twin-beam echo. Since the echo produces two copies of $\Gamma$---one in the signal sector and one in the idler sector [Eqs.~\eqref{eq:signal-echo} and~\eqref{eq:idler-echo}]---combining measurement outcomes improves sensitivity. For instance, suppose we aim to estimate the displacement variance in the $k$th mode $\psi_k\in\Psi$, i.e., the element $\Gamma_{kk}^{[\Psi]}$. [This mode decomposition is especially useful if the displacement field is sparse and approximately diagonal in a known basis.] A straightforward calculation shows that excitation measurements (e.g., photon counting) on the $k$th signal-idler pair yields the classical Fisher information
\begin{equation}
  F_{C}\!\left(\Gamma_{kk}^{[\Psi]}\right)\approx \frac{\cosh^2 r_k + \sinh^2 r_k}{\Gamma_{kk}^{[\Psi]}},
\end{equation}
This saturates the quantum Fisher information for single-mode random-displacement channels~\cite{Wolf2019:MotionalFock,Gorecki2022SpreadChannel,Tsang2023NoiseSpectr,Shi2023DMLimits}, previously shown to be achievable with a two-mode squeezed vacuum~\cite{Shi2023DMLimits} or a Fock state~\cite{Wolf2019:MotionalFock} (see Table~\ref{tab:qfi_loss_gain_imaging} for a compact review). Our framework extends these single-mode results to the multimode, multiparameter setting relevant for imaging, enabling reconstruction of the covariance kernel, $\Gamma$ (cf.~Ref.~\cite{Brady2024CorrNoiseQSN}). Moreover, by choosing a spatially structured basis and mode-resolved detection, SPADE-like strategies can be incorporated to achieve subdiffraction resolution if necessary. Cross-correlations (off-diagonal elements $\Gamma_{kj}^{[\Psi]}$), if present, may be estimated by performing structured interferometry followed by photon counting after the echoes, in accordance with techniques analyzed in Ref.~\cite{Feng2025:PhysLayerPCA}.

\begin{table*}
\centering
\renewcommand{\arraystretch}{2.0}
\setlength{\tabcolsep}{12pt}
\begin{tabular}{c c c}
\hline\hline
\textbf{Estimation Task, Parametrization} & \textbf{Probe} & \textbf{Perturbative QFI, $F_Q(\theta)$} \\
\hline
\multirow{2.25}{*}{Pure loss estimation, $\eta=e^{-\gamma_{\rm loss}}$} 
  & Quantum Optimal & $\displaystyle F_Q(\gamma_{\rm loss}) \approx \frac{N_S}{\gamma_{\rm loss}}$~\cite{Monras2007LossEst,Adesso2009LossEst} \\
  & Coherent State & $\displaystyle F_Q(\gamma_{\rm loss}) \approx (1-\gamma_{\rm loss})N_S$ \\
\hline
\multirow{2.25}{*}{Amplification estimation, $G=e^{\gamma_{\rm amp}}$} 
  & Quantum Optimal & $\displaystyle F_Q(\gamma_{\rm amp}) \approx \frac{1+N_S}{\gamma_{\rm amp}}$~\cite{Aspachs2010:UnruhEst} \\
  & Coherent State & $\displaystyle F_Q(\gamma_{\rm amp}) \approx \frac{1}{\gamma_{\rm amp}}+(1-\gamma_{\rm amp})N_S$ \vspace{.3em} \\
\hline
\multirow{2.25}{*}{Stochastic displacement estimation, $\alpha\sim\mathcal{CN}(0,\gamma_{\rm agn}$)} 
  & Quantum Optimal & $\displaystyle F_Q(\gamma_{\rm agn}) \approx \frac{1+2N_S}{\gamma_{\rm agn}}$~\cite{Wolf2019:MotionalFock,Gorecki2022SpreadChannel,Shi2023DMLimits} \\
  & Coherent State & $\displaystyle F_Q(\gamma_{\rm agn}) \approx \frac{1}{\gamma_{\rm agn}}$ \vspace{.3em} \\
\hline
\multirow{3}{*}{Subdiffraction fluorescence imaging, $\mathfrak{s}=d/\sigma$} 
  & Quantum Optimal & $\displaystyle F_Q(\mathfrak{s}) \approx \frac{\varepsilon(1+N_S)}{4}$ [this work]  \\
  & Vacuum & $\displaystyle F_Q(\mathfrak{s}) \approx \frac{\varepsilon}{4}$~\cite{Tsang2016Superresolution,Lupo2016PRL_SubwaveImaging} \\
  & Coherent State & $\displaystyle F_Q(\mathfrak{s}) \approx \frac{\varepsilon}{4}+\frac{N_S}{64}\left(\varepsilon \mathfrak{s}\right)^2$ \vspace{.3em} 
  \\
\hline
\multirow{2.25}{*}{Subdiffraction absorption imaging, $\mathfrak{s}=d/\sigma$} 
  & Quantum Optimal & $\displaystyle F_Q(\mathfrak{s}) \approx \frac{N_S\gamma_\uparrow}{2}$ [this work] \\
  & Coherent State & $\displaystyle F_Q(\mathfrak{s}) \approx \frac{N_S}{16}\left(\gamma_\uparrow \mathfrak{s}\right)^2$ \vspace{.3em} \\
\hline\hline
\end{tabular}
\caption{
Perturbative quantum Fisher information (QFI) bounds for representative (single-parameter) estimation tasks: bosonic loss and amplification (a.k.a., absorption and incoherent bosonic-mode excitation); stochastic displacements (or additive Gaussian noise); and spatial resolution in subdiffraction fluorescence and absorption imaging.  
$N_S$ is the total mean photon number incident on the sample excluding stored idler photons, $\varepsilon$ the total scene brightness, and $\gamma_\uparrow$ the local point-source absorption rate.  
``Vacuum" refers to passive imaging, e.g., collecting fluorescence without active probing.  
“Quantum Optimal’’ denotes probes that saturate the ultimate quantum limit.  
``Coherent state" serves as a semiclassical baseline for comparison.
For faint scenes ($\varepsilon \ll 1$) and in the subdiffraction regime ($\mathfrak{s}\ll1$), SPADE can circumvent Rayleigh's curse~\cite{Tsang2016Superresolution} in passive fluorescence imaging. In contrast, absorption imaging with classical probes remains fundamentally vexed by Rayleigh's curse,
and quantum probes must be cast to dispel it. 
Observe that coherent-state enhancement in active fluorescence imaging also vanishes as $\mathfrak{s}\rightarrow 0$.
Structured twin-beam echoes (or structured Fock-state probes) and excitation measurements achieve these single-parameter bounds for subdiffraction fluorescence and absorption imaging [this work].
Notable estimation results: bosonic loss~\cite{Monras2007LossEst, Adesso2009LossEst}, including thermal-loss extensions~\cite{Monras2011:ThermLoss, Nair2020:LimitsQuIllum} (with explicit receiver design~\cite{Gong2023:TransmitSens}) and multimode generalizations~\cite{Nair2018:LossEst}; amplifier gain~\cite{Aspachs2010:UnruhEst}; stochastic (or phase-insensitive) displacements~\cite{Wolf2019:MotionalFock,Gorecki2022SpreadChannel,Shi2023DMLimits} (see also Ref.~\cite{Tsang2023NoiseSpectr}) including loss~\cite{Shi2023DMLimits, Gardner2025StochWaveform} and energy-unconstrained limits~\cite{Pirandola2017PRL_AdaptiveLimits}; phase-covariant channels~\cite{Nair2023:CovChSensing}; subdiffraction passive fluorescence imaging~\cite{Tsang2016Superresolution, Lupo2016PRL_SubwaveImaging}. Entries marked [this work] are derived herein. The subdiffraction-imaging results assume two identical point objects with known centroid, a Gaussian PSF, weak absorption or emission, and $\mathfrak{s}\ll1$. See Appendix~\ref{app:sensitivity-limits} for derivations and further discussion. The table is a representative compilation rather than a summary of main results.
}
\label{tab:qfi_loss_gain_imaging}
\end{table*}

\subsection{Structured Fock states}

An alternative to the twin-beam echoes is to prepare structured Fock states, 
\begin{equation}
    |n_k\rangle = \frac{(\hat{\psi}_k^\dagger)^{n_k}}{\sqrt{n_k!}} \ket{0},
\end{equation}
where $\hat{\psi}_k$ is the annihilation operator for the spatial mode $\psi_k\in\Psi$, and perform structured photon-counting measurements in this basis on the collection plane. Using Eq.~\eqref{eq:structured-grand}, the (measurement averaged) output state of the $k$th spatial mode becomes
\begin{multline}
    \rho_k(\Theta) \approx \left(1-n_k\Gamma_{\uparrow, kk}^{[\Psi]}-(n_k+1)\Gamma_{\downarrow, kk}^{[\Psi]}\right)\dyad{n_k} \\
    + n_k  \Gamma_{\uparrow, kk}^{[\Psi]} \dyad{n_k-1} \\ 
    + (n_k+1) \Gamma_{\downarrow, kk}^{[\Psi]}\dyad{n_k+1}.\label{eq:fock-out}
\end{multline}
We write the expression out explicitly in the Fock basis to demonstrate that fluorescence and absorption events are encoded in adjacent $\pm1$ photon-number subspaces from the initial Fock level $n_k$. Photon-number-resolving detection (or at least  ability to distinguish $n_k\pm 1$) thus enables independent reconstruction of the fluorescence and absorption components, analogous to the \mbox{signal/idler} separation of twin-beam echoes.

The decomposition in Eq.~\eqref{eq:fock-out} immediately implies equivalent performance to that of twin-beam echoes for all estimation tasks considered above (given a fixed average number of signal photons), with the correspondences $n_k \leftrightarrow \sinh^2 r_k$ and $n_k+1\leftrightarrow \cosh^2r_k$. The key difference is that the relevant information is now encoded in distinct photon-number subspaces, rather than in distinct signal and idler modes. This expected performance aligns with fundamental single-parameter results for bosonic loss estimation~\cite{Adesso2009LossEst}, gain estimation~\cite{Aspachs2010:UnruhEst}, and stochastic displacement estimation~\cite{Wolf2019:MotionalFock,Gorecki2022SpreadChannel}, where Fock states are known to achieve the quantum Fisher information for a fixed number of average photons (see Table~\ref{tab:qfi_loss_gain_imaging}).

Structured Fock states offer a compact, single-sector alternative to twin-beam echoes. Compared to the latter, they avoid the need for idlers (hence, no entanglement assistance is required) and, in principle, offer better resilience in the presence of idler loss (see Section~\ref{sec:noise-effects}). However, structured Fock states are experimentally challenging to prepare, especially for large $n$, and require photon-number resolution in the structured mode basis.

\subsection{What about single-mode squeezing?}
\label{sec:single-sqz}

Single-mode squeezing has been identified as an optimal probe for certain incoherent parameter estimation tasks, such as sensing stochastic displacements~\cite{Gorecki2022SpreadChannel,Shi2023DMLimits,Tsang2023NoiseSpectr} and loss estimation in the perturbative regime~\cite{Monras2007LossEst}, at least in the absence of Gaussian noise sources~\cite{Shi2023DMLimits,Gardner2025StochWaveform,Shi2025MitRayleigh}. In this context, a single-mode squeezing echo combined with photon counting can achieve quantum-enhanced sensitivity~\cite{Burd2019:SqzEcho,Gorecki2022SpreadChannel,Tsang2023NoiseSpectr}. However, as we elaborate here, this approach is ill-suited to the multiparameter estimation problems considered in this work and, more fundamentally, is fragile to noise for each of the parameter estimation tasks (see Section~\ref{sec:noise-effects}).

In the generalized incoherent imaging problem involving both absorption and fluorescence components, the performance of single-mode squeezing is innately limited. Because squeezing acts only on the signal sector---and due to the continuous-variable nature of squeezing---absorption and fluorescence signatures are mapped to indistinguishable thermal excitations within the same mode. In other words, there exists no mechanism within the single-mode squeezing approach to distinguish these otherwise distinct processes. This contrasts sharply with Fock-state probes or twin-beam echoes, where absorption and fluorescence rest in distinct photon-number subspaces or distinct mode sectors, respectively, enabling simultaneous estimation of both components.

We can formalize this observation. Consider the structured imaging evolution of Eq.~\eqref{eq:structured-grand}, and suppose that we apply single-mode squeezing $\hat{U}\hat{S}$ and its inverse $\hat{U}\hat{S}^\dagger$ (i.e., single-mode squeezing echoes~\cite{Burd2019:SqzEcho}) in the structured mode basis $\Psi=\{\psi_k\}$. Under the squeezing echo, the mode operators transform as $\hat{U}\hat{S}^\dagger \hat{\psi}_k \hat{U}\hat{S} = \cosh(r_k)\hat{\psi}_k + \sinh(r_k)\hat{\psi}_k^\dagger$. Performing independent echoes and excitation measurements (which, as in the twin-beam echo, erases two-photon coherence terms) yields the measurement-averaged output state 
\begin{equation}\label{eq:sms-echo}
    \rho^{\rm sms}\approx \left(1-\Gamma_{\rm sms}\right)\dyad{0}+\sum_{\ell,k}\left(\widetilde{\Gamma}_{\uparrow, k\ell}^{[\Psi]}+\widetilde{\Gamma}_{\downarrow, \ell k}^{[\Psi]}\right)\hat{\psi}_\ell^\dagger\dyad{0}\hat{\psi}_k, 
\end{equation}
with effective rates $\widetilde{\Gamma}_{\uparrow, k\ell}^{[\Psi]}=\Gamma_{\uparrow, k\ell}^{[\Psi]}\sinh(r_k)\sinh(r_\ell)$ and $\widetilde{\Gamma}_{\downarrow, k\ell}^{[\Psi]}=\Gamma_{\downarrow, k\ell}^{[\Psi]}\cosh(r_k)\cosh(r_\ell)$. 
We see that both fluorescence and absorption components contribute to the same single-excitation subspace, preventing simultaneous estimation. 

Although optimal for many incoherent single-parameter estimation tasks in the perturbative regime~\cite{Monras2007LossEst,Aspachs2010:UnruhEst,Gorecki2022SpreadChannel, Tsang2023NoiseSpectr}, single-mode squeezed states prove ill-suited for the multiparameter estimation tasks that we consider. This shortcoming also underlies their inherent fragility to (Gaussian) noise sources. As we discuss in Section~\ref{sec:noise-effects} below, even perturbatively small loss or heating inevitably contaminates the image with spurious excitations, nullifying the quantum advantage that single-mode squeezing might otherwise bestow.

\subsection{Perturbative impact of noise}
\label{sec:noise-effects}

\begin{table*}[t]
\centering
\renewcommand{\arraystretch}{1.5}
\setlength{\tabcolsep}{8.5pt}
\begin{tabular}{c c c c c}
\hline\hline
\textbf{Noise Source} & \textbf{Probe} & \textbf{Absorp.} & \textbf{Fluor.} & \textbf{Comment} \\
\hline
\multirow{3}{*}{Excitation loss} 
  & Twin-beam echo        & \xmark & \cmark & Loss mapped to idler (fluor. preserved in signal) \\
  & Fock                     & \xmark & \cmark & Loss mapped to $n-1$ (fluor. preserved in $n+1$) \\
  & Squeezed vacuum         & \xmark & \xmark & Indistinguishable outputs, signal corrupted \\
\hline
\multirow{3}{*}{Heating} 
  & Twin-beam echo        & \cmark & \xmark & Heating remains on signal (absorp. preserved in idler) \\
  & Fock                     & \cmark & \xmark & Heating mapped to $n+1$ (absorp. preserved in $n-1$) \\
  & Squeezed vacuum                      & \xmark & \xmark & Indistinguishable outputs, signal corrupted \\
\hline
\multirow{3}{*}{Additive noise} 
  & Twin-beam echo        & \xmark & \xmark & Both channels (signal and idler) corrupted \\
  & Fock                     & \xmark & \xmark & Both channels ($n\pm 1$) corrupted \\
   & Squeezed vacuum                     & \xmark & \xmark & Indistinguishable outputs, signal corrupted \\
\hline\hline
\end{tabular}
\caption{Perturbative noise resilience for various quantum probes: Twin-beam echo (clean idler); Fock-state probe; squeezed vacuum (single-mode squeezing of the signal mode). Checkmarks (\cmark{}) indicate robustness to noise while crosses (\xmark{}) indicate fragility to noise in the respective imaging context (Absorp. = absorption or Fluor. = fluorescence). Analysis assumes noise is weak but comparable to (or stronger than) absorption and fluorescent signatures. Twin-beam echoes and Fock states perform identically when idlers are noiseless. The twin-beam-echo entries assume a clean idler channel, for which idler loss and heating are negligible on the scale of the relevant imaging signatures. If both signal and idler modes are sufficiently noisy (dirty idler), then the performance of twin-beam echoes suffers, similar to squeezed vacuum.} 
\label{tab:noise_summary}
\end{table*}

We examine the impact of (Gaussian) noise sources, such as excitation loss, heating, and stray background light (a.k.a., additive Gaussian noise), on imaging performance. We focus on the regime of weak noise, where effects can be treated perturbatively. For instance, suppose that loss and heating take a diagonal form in the mode basis $\Psi$, with $\kappa^{\rm loss}_k$ and $\kappa^{\rm heat}_k$ denoting the rates for the $k$th mode. To first order, loss and heating transforms the absorption and fluorescence profiles [Eq.~\eqref{eq:structured-grand}] as $\Gamma_{\uparrow,\ell k}^{[\Psi]} \rightarrow \Gamma_{\uparrow,\ell k}^{[\Psi]}+\kappa_k^{\rm loss}\delta_{\ell k}$ and $\Gamma_{\downarrow,\ell k}^{[\Psi]} \rightarrow \Gamma_{\downarrow,\ell k}^{[\Psi]}+\kappa_k^{\rm heat}\delta_{\ell k}$. Additive noise affects both channels equally so that formally $\kappa^{\rm agn}_k\coloneqq \kappa_k^{\rm heat}=\kappa_k^{\rm loss}$ above. Hence, decoherence from these mechanisms manifests as spurious backgrounds that must be avoided or at least accounted for (i.e., subtracted off) in post-processing.  

The twin-beam echo provides a clear conceptual basis for understanding how these noise mechanisms affect measurement sensitivity in both absorption and fluorescence imaging, and we elaborate on this in what follows. Similar conclusions hold for Fock-state probes under appropriate conditions, which we discuss below. We also highlight the fragility of single-mode squeezed states. For a summary of these effects across probe types and noise sources, see Table~\ref{tab:noise_summary}.

\paragraph*{\textbf{Signal-mode loss.}} Excitation loss on the signal mode (quantified by $\kappa^{\rm loss}_k$), occurring in the middle of the twin-beam echo, is mapped entirely onto the idler mode by the echo sequence. Therefore, signal-mode loss induces spurious thermal excitations into the idler channel, where absorption features reside, thereby contaminating the absorption image. When the excitation loss becomes comparable to or exceeds the absorption rate, the estimation precision is ultimately limited by the physical signal-to-noise ratio between the two competing processes. On the other hand, the fluorescent component on the signal mode is unaffected by excitation loss to first order. In other words, fluorescence imaging via twin-beam echoes is robust against signal-mode loss.

\paragraph*{\textbf{Signal-mode heating.}} 
Spurious excitations, e.g., heating, on the signal mode (quantified by $\kappa^{\rm heat}_k$) remain localized to the signal mode after the echo. In this case, the absorption features (mapped to the idlers) are protected, while the fluorescence signatures are corrupted. Hence, to first order, absorption imaging is resilient to signal-mode heating, while fluorescence imaging is corrupted by heating.

\paragraph*{\textbf{Background light.}} 
We formally model stray background light that leaks into the signal mode as additive Gaussian noise~\cite{Oh2021NoisySuperres}, corresponding to a stochastic field-displacement channel on the input (quantified by $\kappa^{\rm agn}_k$). This noise channel can be operationally viewed as a balanced combination of excitation loss and heating. Consequently, additive noise ultimately contaminates both fluorescence (signal-mode) and absorption (idler-mode) images. In this case, the imaging problem bears similarities to passive imaging in the presence of thermal backgrounds, where Rayleigh’s curse reemerges and the classical signal-to-noise ratio constrains sensitivity (cf. Refs.~\cite{Oh2021NoisySuperres,Lupo2020:NoisyDetect}). Quantum techniques, such as structured probes and SPADE, nevertheless remain valuable in this setting, but absolute performance is bounded by backgrounds.

\paragraph*{\textbf{Idler noise.}} Due to the signal-idler symmetry of the twin-beam echo, noise on the idler mode degrades both fluorescence and absorption images. Idler-mode loss induces thermal excitations on the signal sector after the echo, corrupting fluorescence imaging. Likewise, idler-mode heating taints the idler mode which is used for absorption imaging. When idler loss becomes comparable to the intrinsic signal brightness $\varepsilon$, the echo-induced amplification actually works against the fluorescence signature---acting as background noise---and restores Rayleigh’s curse (cf.~Refs.~\cite{Gardner2025StochWaveform,Shi2023DMLimits}). Further, if background noise in the idler is comparable to or higher than the signal channel, then the performance of the twin-beam echo protocol degrades.

\paragraph*{\textbf{Comparison with Fock probes.}} Structured Fock states and twin-beam echoes exhibit identical noise resilience to first order when the idler mode is noiseless (see Table~\ref{tab:noise_summary}). Both are robust against excitation loss in fluorescence imaging and heating in absorption imaging. For Fock-state probes, this robustness arises because fluorescence signatures (and heating) are mapped to the $n+1$ excitation subspace, while absorption signatures (and loss) are mapped to the $n-1$ excitation subspace. The key difference between Fock-state probes and twin-beam echoes emerges only when the idler is noisy. In this case, Fock states (and non-Gaussian states more generally) are, in principle, more promising for estimation due to better noise resilience~\cite{Gardner2025StochWaveform}. However, structured Fock states pose practical challenges, including the need for high-fidelity state preparation and photon-number-resolving measurements in a structured mode basis (or, at minimum, the ability to distinguish $n \pm 1$ levels).

\paragraph*{\textbf{Fragility of single-mode squeezing.}} 
As discussed in Section~\ref{sec:single-sqz}, single-mode squeezing echoes cannot distinguish absorption and fluorescence components because both are mapped to the same single-excitation subspace [Eq.~\eqref{eq:sms-echo}]. Consequently, any excitation loss or heating transforms into spurious thermal excitations on the same subspace and inevitably contaminates the image. This vulnerability is not unique to absorption and fluorescence imaging but extends to displacement-field estimation by similar reasoning~\cite{Gardner2025StochWaveform,Shi2023DMLimits,Shi2025MitRayleigh}. In fact, any Gaussian continuous-variable resource appears inadequate for incoherent sensing tasks when spurious noise affects \emph{all} modes~\cite{Gardner2025StochWaveform}; the twin-beam echo is no exception once the idler becomes sufficiently noisy.

\section{Potential Applications}
\label{sec:apps}

The framework developed here has potential applications across diverse imaging and sensing platforms. In this section, we highlight examples where quantum enhancements may be possible. This list is illustrative rather than exhaustive.

\paragraph*{\textbf{Quantum optical microscopy.}}  
Optical microscopy often involves detecting weak signals arising from incoherent processes, such as spontaneous emission, absorption, or scattering, in both the far-field and near-field settings. Quantum advantages in these contexts are most relevant when signals are intrinsically photon-starved, probe photons must be limited to avoid sample damage, or subdiffraction spatial resolution is required~\cite{Taylor2016QuMetrologyBio,Defienne2024AdvQuImaging}.

Quantum-enhanced Raman microscopy, for example, has demonstrated improved signal-to-noise using squeezed light in coherent stimulated setups~\cite{Casacio2021QuNLmicroscoy}. Extending quantum techniques to incoherent (spontaneous) emission could facilitate quantum-enhanced fluorescence microscopy or stimulated imaging without a phase reference. In principle, one could seed structured Stokes modes (and measure in an appropriate basis) to improve sensitivity and resolution in this context.

Photonic force microscopy, where a nanoparticle trapped at the tip of a scanning probe functions as a sensitive mechanical sensor, represents another domain fit for quantum enhancements. Indeed, Taylor \emph{et al.}~\cite{Taylor2014SubDiffQuImaging} demonstrated that interrogating the nanoparticle with squeezed light improves positional tracking in complex biological environments. Applying Fock-state probes or twin-beam echoes and performing photon counting, rather than quadrature readout, could enhance sensitivity to subtle fluctuations or diffusive motion that would otherwise be obscured by vacuum fluctuations in a shot-noise-limited platform.

Quantum advantages have also been demonstrated in absorption microscopy. Spatially correlated twin beams have been deployed to image weak absorbers with reduced noise~\cite{Brida2010SubShotImaging}, and sub-shot-noise wide-field imaging has been demonstrated~\cite{Samantaray2017QuWideMicroscope,Moreau2017:AbsoluteAbsorp}. These strategies align naturally with our developments and offer pathways to further improve both sensitivity and spatial resolution (a.k.a., subdiffraction absorption imaging).

\paragraph*{\textbf{Quantum plasmonic imaging.}}

Plasmonic sensing represents a promising frontier for quantum-enhanced imaging. In these systems, collective electron oscillations in nanoscale metallic structures amplify local electromagnetic fields and enhance light–matter interactions to improve sensing. Recent demonstrations with quantum probes include the use of twin-beam states for improved refractive-index sensitivity in plasmonic nanostructures~\cite{Dowran2018:QuPlasmonSens} and entangled-photon spectroscopy for detecting plasmonic resonance shifts under noise~\cite{Kalashnikov2014:PlasmonicNano}; see Ref.~\cite{Lee2021:QuPlasRvw} for a review. While these efforts have focused on spectroscopy, our framework facilitates \emph{spatial} plasmonic imaging. This opens the possibility for real-time mapping of nanoscale absorption, fluorescence features, and near-field distributions---effectively forming the basis of a quantum plasmonic \emph{microscopy} platform. Tailoring multiple degrees of freedom in tandem (viz., spatiotemporal mode shaping) could allow joint spatial and spectral imaging of plasmonic systems, with applications in ultrathin materials, nanophotonic metasurfaces, or bio-sensing platforms. An interesting direction along these lines is direct plasmonic state engineering, enabling tailored quantum–plasmonic resources for imaging and sensing.

\paragraph*{\textbf{Wide-field optical readout.}}
Our methods may benefit advanced (wide-field) quantum optical readout of quantum sensors, such as nitrogen-vacancy (NV) center arrays in diamond~\cite{Levine2019nvScope,Scholten2021WideNVmicroscope,Deshler2025:nvSPADE}. These systems often rely on fluorescence detection to infer spin-dependent signals---which may be stochastic due to the chaotic nature of nearby electromagnetic fields---but face limitations in sensitivity and resolution under photon-starved conditions. Quantum probes could amplify weak readout signals, particularly for multiplexed optical detection~\cite{Cheng2025:nvMultiplex,Cambria2025:nvMultiplex, Deshler2025:nvSPADE}. In principle, these techniques may improve dynamic field mapping, correlated fluorescence imaging~\cite{Ji2024nvRandEfield}, and exploration of many-body phenomena in condensed-matter systems~\cite{Casola2018nvCM,Rovny2024nvManyBody, Nambiar2025:PhasesViaPhotonCorr}.

\paragraph*{\textbf{Phononic imaging.}}
Quantum imaging of mechanical vibrations, whether in optomechanical membranes or solid-state phononic systems, provides a natural playground for our imaging protocols. 

Recent experiments have demonstrated quantum-limited readout of (coherent) mechanical motion at spatially distinct points on a single mechanical membrane. Choi \emph{et al.}~\cite{Choi2025OmechImaging} imaged nanoribbon vibrations using a multi-plane light conversion (MPLC) mode sorter and homodyne detection, achieving near-standard quantum limit displacement sensitivity. Pluchar \emph{et al.}~\cite{Pluchar2025ActiveQuImaging_Omech} provided the theoretical basis, showing that structured quadrature measurements can resolve flexural membrane modes.

These demonstrations target \emph{coherent} displacement estimation, where homodyne detection is well suited. By contrast, the framework developed here applies to \emph{incoherent} phononic imaging, where thermally driven fluctuations or stochastic displacements comprise the signal. In which case excitation measurements become essential because quadrature detection is limited by vacuum noise. Our methods could enable quantum-enhanced imaging of flexural mode fluctuations below the vacuum noise floor.

Beyond optical interrogation, quantum control of phonons in circuit quantum acoustodynamics (cQAD) allows direct preparation and measurement of mechanical excitations. Recent demonstrations of Fock-state preparation and parity-sensitive phonon readout~\cite{Schrinski2024Fadel-acoustic,Rahman2025:CQADfock} highlight the potential for metrological applications. Extending these ideas to multimode phononic imaging could allow spatial reconstruction of dissipation landscapes or improved monitoring of decoherence dynamics.

\paragraph*{\textbf{Distributed harmonic oscillators.}} 
More broadly, distributed arrays of coupled quantum harmonic oscillators---including trapped ions, optomechanical sensors, and microwave cavities---offer versatile platforms for imaging and sensing weak, spatially varying stochastic forces and fields.

In trapped ion systems, motional fluctuations can encode information about electric fields, fluctuating patch potentials, or other correlated noise sources~\cite{Brownnutt2015:IonEnoise}. Quantum probes based on motional Fock states~\cite{Wolf2019:MotionalFock} and squeezed states~\cite{Burd2019:SqzEcho} have demonstrated enhanced displacement sensitivity in highly controlled single-ion systems, while squeezed states of collective modes in an ion crystal have enabled distributed electric-field sensing~\cite{Gilmore2021IonEFieldQSN}.

Similar opportunities exist in optomechanical sensor networks, which have been proposed and implemented for distributed force sensing~\cite{Brady2023OmechArray,Xia2023OmechDQS} and radiofrequency field detection~\cite{Xia2020:rfField,Xia2021:SLAEN}. Current demonstrations typically rely on single-mode squeezing and homodyne detection. However, a distributed Fock-state probe or twin-beam echo, along with photon-counting at the measurement end, could substantially improve sensitivity to stochastic background fields and enable multiplexed reconstruction of noise patterns across the array in the shot-noise dominant regime.

In the microwave domain, arrays of high-Q (e.g., superconducting) cavities have been considered for detecting weak spatially-correlated fields, including dark-matter search with homodyne detetion of distributed squeezed vacuum~\cite{Brady2022:DMsearch}. Recent advances have demonstrated the use of Fock-state probes combined with parity detection for displacement sensing in a single cavity~\cite{Agrawal2024:FockDM}, again targeting new-physics searches. It is tantalizing to envision extensions to a multi-cavity sensor network.

\section{Outlook}
\label{sec:outlook}

We introduce a quantum universal imaging module capable of simultaneously imaging absorption and fluorescence profiles as well as reconstructing quadrature-displacement fields in the perturbative regime of weak signals. This unified approach achieves quantum-enhanced sensitivity and subdiffraction resolution (where applicable), all within a common framework.

As we show in this work, twin-beam echoes provide a physically motivated mechanism for isolating and amplifying signal components in distinct modes, while Fock states offer a path to high-precision sensing within distinct photon-number subspaces without the need for ancillary entanglement. The choice between these approaches depends on experimental factors such as mode-matching precision, system imperfections, and the tradeoff between complexity and desirable quantum advantage.

That said, quantum benefits do not come for free. Performance is ultimately constrained by dominant noise mechanisms, with excitation loss, mode mismatch, parasitic heating, and spurious backgrounds being major culprits. We expect the largest gains in shot-noise-limited platforms featuring high-efficiency detection and, perhaps, pristine quantum control---such as cavity-enhanced architectures, waveguide-integrated sensors~\cite{Sheremet2023wqed,Reitz2022cooperative}, and nanophotonic devices that support coherent light-matter interactions and near-unity collection efficiency~\cite{Gonzalez2024:LightMatter,Lednev2025:ResolvedPhotStat}. Still, meaningful enhancements should remain accessible with more modest resources, especially in signal-starved, timing-constrained settings where photon efficiency is paramount.

Several directions remain. A more comprehensive analysis of sensor platforms is warranted---for instance, in phononic or acoustic imaging using optomechanical membranes~\cite{Pluchar2025ActiveQuImaging_Omech,Choi2025OmechImaging}, which requires modeling mechanical fluctuations and radiation-pressure in greater detail, or in reconstructing electric fields and charge distributions using ion crystals or trapped-ion arrays~\cite{Burd2019:SqzEcho,Gilmore2021IonEFieldQSN,Brownnutt2015:IonEnoise}. Extensions to more realistic models, including lossy propagation, and complicated scenes in quantum-optical microscopy also merit systematic study. Another pressing challenge is to extend these ideas to learning structured functionals of the full correlation matrices (the $\Gamma$'s), especially in scenarios with slow dynamics or temporally varying correlations. Examples include extracting spatial correlation lengths, collective parameters, or tracking dynamical observables encoded in the emitters’ two-point functions. Developing optimal quantum imaging protocols for task-specific objectives remains an outstanding problem.

Regarding our theoretical approach, our framework has been developed within the perturbative regime of single-photon dynamics. Future work should move beyond this limit to investigate multi-photon processes, including a more complete analysis of noise effects. At elevated noise levels, connections to quantum illumination~\cite{Tan2008:quIllumination,Guha2009:quIllumination,Zhuang2017:quIllumination} become relevant, especially for cluttered scenes~\cite{Cox2024:quIllumination}. The circuit-level modeling that we employ should also be extended to the continuous-time setting for dynamical sensing and imaging, where rapid quantum control may offer further advantages in principle~\cite{Sekatski2017:FastQuCtrl,Gardner2025LindbladEst}. Likewise, variational methods for probe and measurement design---perhaps leveraging physical-layer principal component analysis~\cite{Feng2025:PhysLayerPCA}---might prove effective for complex-scene interrogation.

While our work focuses on spatial imaging, the underlying formalism should be extendable to joint spatio-spectral domains. By engineering both the transverse spatial structure and the spectral (or temporal) profile of the probe field, one could target the frequency-dependence of absorption or fluorescence processes, enabling quantum-enhanced spectroscopy atop improved spatial imaging. This would require adopting tensor-product mode structures~\cite{Huang2023ExoSpectroscopy} and co-designing probe generation and measurement strategies accordingly. Concurrently, extending the framework to full three-dimensional imaging, capturing both lateral and longitudinal structure, remains an open avenue.

Recent insights from the literature provide further motivation and broader conceptual context of our work. Tsang~\cite{Tsang2023NoiseSpectr} recently illuminated an intimate connection between incoherent optical imaging and stochastic waveform estimation. Our results reinforce and extend this perspective to a multi-modal paradigm that directly addresses displacement-field reconstruction and incoherent imaging in tandem. In another light, Albarelli \emph{et al.}~\cite{Albarelli2020Perspective} have suggested that: ``Further insight could possibly be gained by casting quantum imaging as a problem in distributed sensing." Our structured quantum-probe framework, while not phrased in those terms, emulates a distributed quantum state across spatial degrees of freedom of the probe field, carrying the same structural seeds of distributed quantum sensing~\cite{Zhuang2018PRA_dqs, Ge2018PRL_qsn, Guo2020Nat_DQSphase, Zhang2021DQSrvw, Bringewatt2024} in the context of correlated noise estimation~\cite{Brady2024CorrNoiseQSN}. More fruitful connections between quantum imaging and distributed quantum sensing are bound to ripen.

The quantum imaging protocols developed here add another layer of functionality and breadth to the already rich landscape of multiparameter quantum imaging. They are adaptable to a wide range of platforms and modalities, with the potential to surpass classical resolution limits and, in parallel, achieve quantum-enhanced measurement sensitivity using current- or next-generation quantum imaging technology.

\acknowledgments
A.J.B. acknowledges Wenhua He, Haowei Shi, and Yuxin Wang for valuable conversations.
A.J.B.\@ acknowledges support from the NRC Research Associateship Program at NIST.
Saikat Guha and Zihao Gong acknowledge the support by the Office of Naval Research (ONR) under grant number N00014-19-1-2189.
A.V.G.~was supported in part by ARL (W911NF-24-2-0107), AFOSR MURI, ONR MURI, DARPA SAVaNT ADVENT, DoE ASCR Quantum Testbed Pathfinder program (awards No.~DE-SC0019040 and No.~DE-SC0024220), NSF QLCI (award No.~OMA-2120757), NSF STAQ program, and NQVL:QSTD:Design:FTL. A.V.G.~also acknowledges support from the U.S.~Department of Energy, Office of Science, National Quantum Information Science Research Centers, Quantum Systems Accelerator (QSA) and from the U.S.~Department of Energy, Office of Science, Accelerated Research in Quantum Computing, Fundamental Algorithmic Research toward Quantum Utility (FAR-Qu).

\bibliography{main}

\onecolumngrid
\appendix

\section{Constructing an Incoherent Imaging Model}
\label{app:model}

In this appendix, we develop our theoretical imaging model based on a toy microscopic description of the (local) emitter-probe interaction at the emitter plane, followed by free-space propagation to the collection plane. The probe is a quantum optical field described in terms of spatial modes, which we model as a quantum oscillator at each position (see Ref.~\cite{Fabre2020QuOpticsReview} for background on quantum optics). We suppose that all the emitters lie on a single transverse plane (say at $z=0$) and, thus, focus exclusively on the transverse spatial degree of freedom of the probe field. For simplicity, we restrict to one transverse spatial dimension and a single carrier frequency and neglect polarization, which would accompany the analysis without qualitatively altering the results. 

We operate in the weak-interaction regime, such that the probe does not significantly influence the internal dynamics of the emitters. Physically, this corresponds to scenarios where the emitters are separately driven by their environment, and the probe serves only as a minimally invasive diagnostic. By the same token, single-photon processes dominate the probe evolution while multi-photon events are suppressed. We further assume that the emitter dynamics are sufficiently rapid, so that they behave as a Markovian bath from the viewpoint of the probe. This description captures the essential features of quantum imaging in a simplified context. See below for further details.

\subsection{Emitter-probe interaction} 
\label{app:emitter-probe}

We consider a quantum-optical pulse of duration $T$ interacting with a Markovian bath of emitters from time $t=0$ to $t=T$. The pulse is sufficiently long so that it may be treated, to a good approximation, as a longitudinal plane wave with a non-trivial transverse profile (i.e., an ideal paraxial beam): This implies that $1/T \ll \omega_0$, where $\omega_0$ is the carrier frequency, so that the slowly-varying-envelope approximation holds. We also assume $T$ is much longer than the coherence time of the emitter bath to ensure strict Markovianity of the emitters. Throughout, we work in the rotating (or propagating) frame of the pulse, that is, with respect to the longitudinal plane wave $e^{i(k_0 z-\omega_0 t)}$.

We assume spatially local interactions between the probe field and emitters on the emitter plane (coordinate $x$). We take the initial emitter-probe state prior to interaction to be separable, $\rho_{S}\otimes\tau_E$, where $\rho_S$ ($S$ for signal modes) denotes the initial state of the probes at $t=0$ and $\tau_E$ the initial state of the emitters. The joint state after interaction is 
\begin{equation}
    \rho_{S E}(\Theta)=\hat{U}_{SE}(T)(\rho_{S }\otimes\tau_E)\hat{U}_{SE}^\dagger(T),
\end{equation}
where $\Theta$ represents a collection of parameters encoded into the probe, and the (time ordered) interaction unitary is 
\begin{equation}
    \hat{U}_{SE}(T)=\mathcal{T}\exp[-i\int_0^T\dd{t}\hat{H}_{SE}(t)]
\end{equation}
with $\hat{H}_{SE}(t)$ the time-dependent interaction operator. We model the coupling as a spatially-local exchange interaction, 
\begin{equation}
    \hat{H}_{SE}(t)=\int\dd{x}g_x(\hat{\mathfrak{a}}_x\hat{\mathfrak{e}}_x^\dagger(t) + \hat{\mathfrak{a}}_x^\dagger\hat{\mathfrak{e}}_x(t)),
\end{equation}
where $g_x\in\mathbb{R}$ characterizes the interaction strength at point $x$. The operator $\hat{\mathfrak{a}}_x$ denotes the time-independent (in the rotating frame) annihilation operator for the probe field on the emitter plane, which obeys $\comm*{\hat{\mathfrak{a}}_x}{\hat{\mathfrak{a}}_{x'}^\dagger}=\delta(x-x')$, and $\hat{\mathfrak{e}}_x(t)$ denotes the Markovian emitter operators such that, e.g., $\expval*{\hat{\mathfrak{e}}_x(\tau)\hat{\mathfrak{e}}^\dagger_x(0)}\propto \delta(\tau)$, conveying Markovianity. Note that we model the emitters in the continuum (i.e., as a dense set) for simplicity, though analyzing a finite set with discrete mode and emitter labels is straightforward (see examples below and, e.g., Ref.~\cite{Tsang2016Superresolution} for related discussion). See Fig.~\ref{fig:interaction-schematic} of the main text for a quantum-circuit representation. In what follows, we consider two different physical realizations (oscillators and two-level systems) for the emitters which determine $\hat{\mathfrak{e}}_x$.

\paragraph*{\textbf{Oscillator emitters.}} 
Here we model the emitters as oscillators. Define the local interaction operator $\hat{\mathfrak{h}}_x=\hat{\mathfrak{a}}_x\hat{\mathfrak{e}}_x^\dagger + \hat{\mathfrak{a}}_x^\dagger\hat{\mathfrak{e}}_x$ and assume weak coupling (i.e., small $g_x$---we give the precise condition below). We expand $\hat{U}_{SE}(T)$ in a Dyson series,
\begin{equation}
    \hat{U}_{SE}(T)\approx \hat{I} - i\int_{0}^T\dd{t}\int\dd{x}g_x\hat{\mathfrak{h}}_x(t) -\int_{0}^T\dd{t}\int_{0}^t\dd{t'}\int\dd{x}\dd{x'}g_xg_{x'}\hat{\mathfrak{h}}_x(t)\hat{\mathfrak{h}}_{x'}(t'),
\end{equation}
with the spatial integral corresponding to a sum over the emitter plane. From which we obtain an expression for the joint state post-interaction:
\begin{multline}
    \rho_{SE}(\Theta)\approx \rho_S\otimes\tau_E - i\int_{0}^T\dd{t}\int\dd{x}g_x\comm{\hat{\mathfrak{h}}_x(t)}{\rho_{S}\otimes\tau_E} \\ + \int_{0}^T\dd{t}\int_{0}^t\dd{t'}\int\dd{x}\dd{x'}g_xg_{x'}\left(\hat{\mathfrak{h}}_x(t)\rho_{S}\otimes\tau_E\hat{\mathfrak{h}}_{x'}(t')-\frac{1}{2}\acomm{\hat{\mathfrak{h}}_x(t)\hat{\mathfrak{h}}_{x'}(t')}{\rho_{S}\otimes\tau_E}\right),
\end{multline}
where $\rho_S\otimes \tau_E$ represents the initial state.

We only access the emitters indirectly by measurements on the probes. Thus, we must trace over the emitter degrees of freedom and find an expression solely for the probe output, $\rho_{S}(\Theta)=\Tr_E\{\rho_{SE}(\Theta)\}$. We assume that the emitters are spatially uncorrelated with vanishing first moments, implying that $\Tr\{\tau_E\hat{\mathfrak{e}}_x\}=\Tr\{\tau_E\hat{\mathfrak{e}}_x\hat{\mathfrak{e}}_{x'}\}=0$, $\Tr\{\tau_E\hat{\mathfrak{e}}_x(\tau)\hat{\mathfrak{e}}_{x'}^\dagger(0)\}=(\bar{n}_x+1)\delta(x-x')\delta(\tau)$, and $\Tr\{\tau_E\hat{\mathfrak{e}}_x^\dagger(\tau)\hat{\mathfrak{e}}_{x'}(0)\}=\bar{n}_x\delta(x-x')\delta(\tau)$, where $\bar{n}_x$ is the emitter population per point $x$. [We extend this model to include emitter-emitter spatial correlations below.] The following relations then hold:
\begin{align}
    \Tr_E\left(\comm{\hat{\mathfrak{h}}_x}{\rho_{S}\otimes\tau_E}\right)&=0,\label{eq:nodrive}\\
    \Tr_E\left(\hat{\mathfrak{h}}_x(t)\rho_{S}\otimes\tau_E\hat{\mathfrak{h}}_{x'}(t')\right)&=\Big(\hat{\mathfrak{a}}_x\rho_{S}\hat{\mathfrak{a}}_x^\dagger(\bar{n}_x+1)+\hat{\mathfrak{a}}_x^\dagger\rho_{S}\hat{\mathfrak{a}}_x\bar{n}_x\Big)\delta(x-x')\delta(t-t'),\\
    \Tr_E\left(\hat{\mathfrak{h}}_x(t)\hat{\mathfrak{h}}_{x'}(t')\rho_{S}\otimes\tau_E\right)&=\left(\hat{\mathfrak{a}}_x\hat{\mathfrak{a}}_x^\dagger\rho_{S}\bar{n}_x+\hat{\mathfrak{a}}_x^\dagger\hat{\mathfrak{a}}_x\rho_{S}(\bar{n}_x+1)\right)\delta(x-x')\delta(t-t').
\end{align}
Note that the first equality implies that the emitters do not coherently drive the probe field, while the second and third equalities follow from Markovianity of the emitters. Given that $\hat{\mathfrak{a}}_x$ are time-independent operators, we deduce, after a trivial time integral, the approximate form of the output probe state as seen on the emitter plane:
\begin{align}
    \rho_{S}(\Theta) \approx \rho_{S} &+  T\int\dd{x}\Bigg[\,\overbrace{g_x^2(\bar{n}_x+1)\left(\hat{\mathfrak{a}}_x\rho_{S}\hat{\mathfrak{a}}_x^\dagger-\frac{1}{2}\acomm{\hat{\mathfrak{a}}_x^\dagger\hat{\mathfrak{a}}_x}{\rho_{S}}\right)}^{\text{Absorption}}+\overbrace{g_x^2\bar{n}_x\left(\hat{\mathfrak{a}}_x^\dagger\rho_{S}\hat{\mathfrak{a}}_x-\frac{1}{2}\acomm{\hat{\mathfrak{a}}_x\hat{\mathfrak{a}}_x^\dagger}{\rho_{S}}\right)}^{\text{Incoherent emission (a.k.a., fluorescence)}}\,\Bigg].\label{eq:rho_oscillator}
\end{align}
Defining the absorption and decay rates, $\gamma_{\uparrow}(x)=g_x^2T(\bar{n}_x+1)$ and $\gamma_{\downarrow}(x)=g_x^2T\bar{n}_x$, we recover Eq.~\eqref{eq:grandfather} of the main text. In this regime, single-photon processes (i.e., single jumps) govern the probe evolution. Also, observe that the imprint of the emitter interaction accumulates linearly with the pulse duration $T$, consistent with Markovian evolution (cf.~Ref.~\cite{Albarelli2023:PulsedEst}); accordingly, we take $T=1$ throughout for convenience. We see that the perturbative expansion is valid provided that $\int\dd{x}\gamma_{\uparrow}(x)\Tr{\rho_S\hat{\mathfrak{a}}_x^\dagger\hat{\mathfrak{a}}_x},\,\int\dd{x}\gamma_{\downarrow}(x)\Tr{\rho_S\hat{\mathfrak{a}}_x\hat{\mathfrak{a}}_x^\dagger} \ll 1$. In other words, the probe energy (per point $x$ and over duration $T$) must not be too large.

\paragraph*{\textbf{Two-level emitters.}}
Here we model the emitters as dense set of two-level systems such that $\hat{\mathfrak{e}}_x=\hat{\sigma}_{-}^{(x)}$ and $\hat{\mathfrak{e}}_x^\dagger=\hat{\sigma}_{+}^{(x)}$ where, $\hat{\sigma}_-=\dyad{0}{1}$ induces a transition from the excited $\ket{1}$ state to the ground state $\ket{0}$. Assuming spatio-temporally uncorrelated emitters, the two-point are correlators $\Tr\{\tau_E\hat{\sigma}_{-}^{(x)}(t)\hat{\sigma}_{+}^{(x')}(t')\}=p_{0}(x)\delta(x-x')\delta(t-t')$ and $\Tr\{\tau_E\hat{\sigma}_{+}^{(x)}(t)\hat{\sigma}_{-}^{(x')}(t')\}=p_{1}(x)\delta(x-x')\delta(t-t')$ where $p_{0}$ ($p_1$) represents the ground (excited) state population density of the emitters. Going through similar calculations as above, we deduce the output probe state:
\begin{align}
    \rho_{S}(\Theta) \approx \rho_{S} +  \int\dd{x}\Bigg[\,\overbrace{g_x^2p_0(x)\left(\hat{\mathfrak{a}}_x\rho_{S}\hat{\mathfrak{a}}_x^\dagger-\frac{1}{2}\acomm{\hat{\mathfrak{a}}_x^\dagger\hat{\mathfrak{a}}_x}{\rho_{S}}\right)}^{\text{Absorption}} +\overbrace{g_x^2p_1(x)\left(\hat{\mathfrak{a}}_x^\dagger\rho_{S}\hat{\mathfrak{a}}_x-\frac{1}{2}\acomm{\hat{\mathfrak{a}}_x\hat{\mathfrak{a}}_x^\dagger}{\rho_{S}}\right)}^{\text{Incoherent emission (a.k.a., fluorescence)}}\,\Bigg],\label{eq:rho_2lvl}
\end{align}
where we take the liberty of setting $T=1$. Recognizing the local absorption and emission rates, $\gamma_{\uparrow}(x)=g_x^2p_0(x)$ and $\gamma_{\downarrow}(x)=g_x^2p_1(x)$, in direct analogy to the oscillator model, we produce Eq.~\eqref{eq:grandfather} of the main text.

\paragraph*{\textbf{Spatially correlated emitters.}}
More generally, suppose that there exist spatial correlations between the emitters, either because of emitter-emitter coupling or non-trivial behavior in the emitters' environment, such that $\Tr\{\tau_E\hat{\mathfrak{e}}_{x}\hat{\mathfrak{e}}_{x'}^\dagger\}=\mathfrak{E}_{\uparrow}(x,x')\delta(t-t')$ and $\Tr\{\tau_E\hat{\mathfrak{e}}_{x'}^\dagger\hat{\mathfrak{e}}_{x}\}=\mathfrak{E}_{\downarrow}(x,x')$ with all other second moments vanishing. These correlations map onto the probes through the emitter-probe interactions via $\gamma_{\uparrow/\downarrow}(x,x')= g_xg_{x'}\mathfrak{E}_{\uparrow/\downarrow}(x,x')$. Including the emitter-emitter correlations, the output state of the probes is  
\begin{align}
    \rho_{S}(\Theta) \approx \rho_{S} +  \int\dd{x}\dd{x'}\left[\gamma_{\uparrow}(x,x')\left(\hat{\mathfrak{a}}_x\rho_{S}\hat{\mathfrak{a}}_{x'}^\dagger-\frac{1}{2}\acomm{\hat{\mathfrak{a}}_{x'}^\dagger\hat{\mathfrak{a}}_x}{\rho_{S}}\right) +\gamma_{\downarrow}(x,x')\left(\hat{\mathfrak{a}}_{x'}^\dagger\rho_{S}\hat{\mathfrak{a}}_x-\frac{1}{2}\acomm{\hat{\mathfrak{a}}_x\hat{\mathfrak{a}}_{x'}^\dagger}{\rho_{S}}\right)\right],
    \label{eq:rho_corr}
\end{align}
which applies to both the oscillator model and two-level systems. A simple example is two positively-correlated point-source emitters of equal brightness, 
\begin{equation}
    \gamma_{\downarrow}(x,x')=\frac{\varepsilon\Delta x}{2}\bigg(\delta(x-x')f(x-x_1)+\delta(x-x')f(x-x_2)+C(d/\xi)\Big(f(x-x_1)f(x'-x_2)+f(x-x_2)f(x'-x_1)\Big)\bigg).
\end{equation}
Here, $\varepsilon$ is the total brightness, $\Delta x$ characterizes the finite emitter size, $f(x-x_i)$ is the spatial density profile for the $i$th emitter, $d=|x_1-x_2|$ the emitter separation, $\xi$ the correlation length, and $0\leq C(d/\xi)\leq 1$ the correlation coefficient. Correlations such as these have been assessed in the context of subdiffraction imaging~\cite{Larson:18correlation}.

\begin{figure}
    \centering
    \includegraphics[width=.45\linewidth]{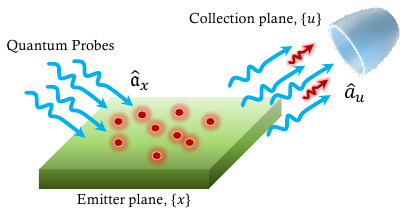}
    \caption{Probe propagation and interaction. Probe light propagates towards the emitter plane, reflects off and propagates towards the collection plane. Outgoing probes carry fluorescence and absorption signatures of the emitters.}
    \label{fig:propagation}
\end{figure}

\subsection{Propagation to collection plane}
Let $u$ denote the collection-plane coordinate and $\hat{a}_u$ the annihilation operator of the probe's pixel mode at $u$ (see Fig.~\ref{fig:propagation}). The collection plane and emitter plane operators, $\hat{a}_u$ and $\hat{\mathfrak{a}}_x$, are related through the (unitary) propagation kernel via
\begin{equation}\label{eq:Kxu}
    \hat{\mathfrak{a}}_x=\int\dd{u}K(x,u)\hat{a}_u,
\end{equation}
where $\comm*{\hat{a}_u}{\hat{a}_{u'}^\dagger}=\delta(u-u')$, and the propagator obeys $\int\dd{u}K(x,u)K^*(x',u)=\delta(x-x')$. The inverse relation holds for forward propagation: 
\begin{equation}\label{eq:Kux}
    \hat{a}_u=\int\dd{x}K^*(x, u)\hat{\mathfrak{a}}_x,
\end{equation}
where $K^*(x,u)$ is the forward propagator. Consequently, the relevant rates at the collection plane are
\begin{align}
    \Gamma_{\uparrow}(u,u')&=\int\dd{x}\dd{x'}\gamma_{\uparrow}(x,x')K(x, u)K^*(x', u'), \label{eq:abs-kern}\\ 
    \Gamma_{\downarrow}(u,u')&=\int\dd{x}\dd{x'}\gamma_{\downarrow}(x,x')K(x, u)K^*(x', u').\label{eq:mut-kern}
\end{align}
where $\Gamma_{\uparrow}(u,u')$ denotes the absorption kernel and $\Gamma_{\downarrow}(u,u')$ the mutual coherence.

Although the propagator $K(x, u)$ is a unitary kernel, it is often replaced by the optical system's point-spread function (PSF) in practice. For example, after spatial filtering by a lens system or single-mode fiber, or taking a Gaussian approximation for $K(x,u)$, one sets $K(x, u)\rightarrow \varphi(u-x)$, where $\varphi(u-x)$ is the normalized PSF centered at $x$. In this context, the rates $\Gamma_{\uparrow/\downarrow}$ should be regarded as physically measured rates. A common approximation is to take $\varphi(u-x)$ as a Gaussian. Physically, the lens system or fiber projects the field onto an (approximate) Gaussian mode of width $\sigma\approx \lambda f/\pi w$, where $f$ is the focal length,  $w$ is the lens aperture radius, and $\lambda$ is the optical wavelength. The Gaussian approximation is well-justified in several contexts relevant to quantum optical imaging, such as imaging point emitters separated by distance $d$ in the far-field regime, $d^2 / (\lambda L) \ll 1$ (the Fraunhofer condition), where $L$ is the propagation distance from the emitter plane to the collection plane, or imaging in the subdiffraction regime, $d / \sigma \ll 1$.

Substituting relations~\eqref{eq:abs-kern} and~\eqref{eq:mut-kern} into Eq.~\eqref{eq:rho_corr}, we find an expression for the output probe on the collection plane:
\begin{equation}\label{eq:app-grand}
    \rho(\Theta)\approx \rho +  \int\dd{u}\dd{u'}\left[\Gamma_{\uparrow}(u,u')\left(\hat{a}_u\rho\hat{a}_{u'}^\dagger-\frac{1}{2}\acomm{\hat{a}_{u'}^\dagger\hat{a}_u}{\rho}\right) +\Gamma_{\downarrow}(u,u')\left(\hat{a}_{u'}^\dagger\rho\hat{a}_u-\frac{1}{2}\acomm{\hat{a}_u\hat{a}_{u'}^\dagger}{\rho}\right)\right],
\end{equation} 
which is the central equation that we analyze throughout our work. 

Finally, note that measurements are performed on the collection plane, while the probe state is initially prepared at a separate source plane. Generally, one must account for propagation from the probe source to the emitter plane (through the kernel $\widetilde{K}$) and from the emitter plane to the collection plane (via $K$) (see Fig.~\ref{fig:interaction-schematic} of the main text). In this work, we assume ideal (lossless) propagation for simplicity.

\section{Subdiffraction Imaging and Rayleigh's Curse: A Brief Overview}
\label{app:imaging}

In this appendix, we briefly review key concepts in incoherent subdiffraction imaging, focusing on the passive single-photon regime. This setting isolates the fundamental problems in imaging, arising with or without quantum probes, and provides a clean context for introducing sophisticated techniques, e.g., SPADE, for imaging beyond Rayleigh's limit. For further background, we refer the reader to Ref.~\cite{Tsang2016Superresolution} and Ch.~13 of Ref.~\cite{Tsang2024notes}.

In the passive case, the input probe state is vacuum (zero probe-photons). Fluorescence from the emissive scene is then modeled as a single-photon incoherent field with mutual coherence matrix $\Gamma(u,u')$. From Eq.~\eqref{eq:app-grand}, the resulting fluorescence state reads
\begin{equation}\label{eq:app-passive-rho}
    \rho(\Theta)\approx (1-\varepsilon)\dyad{0} + \int\dd{u}\dd{u'}\Gamma(u,u')\hat{a}_{u'}^\dagger\dyad{0}\hat{a}_u,
\end{equation}
where $\varepsilon=\int\dd{u}\,\Gamma(u,u)$ denotes the total measured brightness (mean photon number) and $\dyad{0}$ is the vacuum state of the probe field. Since absorption is not analyzed in this appendix, we omit the down-arrow subscript that typically labels emission.

A pertinent example is $M$ uncorrelated point-source emitters. In this case, the mutual coherence takes the form
\begin{equation}\label{eq:app-gamma-uncorr}
    \Gamma(u,u') = \frac{\varepsilon}{M} \sum_{m=1}^M \varphi(u-x_m) \varphi(u'-x_m),
\end{equation}
where $\varepsilon$ represents the total brightness from all emitters, $x_m$ is the position of the $m$th emitter, and $\varphi(u-x)$ is the normalized PSF at the collection plane. Formally, this relation can be derived by assuming the emitters have some very small spatial width, $\Delta x$, that appears in the emission correlator $\gamma_{\downarrow}(x,x')=\varepsilon\Delta x\,\delta(x-x')F(x)$. Here, $\delta(x-x')$ reflects the uncorrelated nature of the emitters while $F(x)=\sum_{m=1}^M f_m(x)$ represents the density profile for the discrete set with $f_m(x)$ being local densities, e.g., a box function $f_m(x)={\rm rect}[(x-x_m)/\Delta x]/\Delta x$. If $\Delta x \ll \sigma$, then $\varphi(u-x)$ varies negligibly over each $f_m(x)$, and we ultimately arrive at Eq.~\eqref{eq:app-gamma-uncorr} via Eq.~\eqref{eq:mut-kern}.

Often, the PSF is approximated as a Gaussian:
\begin{equation}\label{eq:beam_psf}
    \varphi(u) = \frac{e^{-u^2 / 4\sigma^2}}{(2\pi \sigma^2)^{1/4}}.
\end{equation}
An example from optical imaging is $\sigma^2 = w^2(z_0)/4$, where $w(z_0) = w_0 \sqrt{1 + (z_0 / z_R)^2}$ is the Gaussian beam waist on the collection plane, $z_0$ is a reference length, $z_R = \pi w_0^2 /\lambda$ is the Rayleigh length~\cite{Zhou2019zLocalization}, and $\lambda$ the optical wavelength. An important regime is $z_0 / z_R \ll 1$, where $w^2(z_0) \approx z_R \lambda / \pi$. High-resolution imaging at length scales $d \ll \sigma$ corresponds to subdiffraction imaging. At such scales, conventional detection strategies suffer from Rayleigh’s curse, which severely degrades resolution~\cite{Tsang2016Superresolution}. Overcoming this limitation generally requires nonlocal and non-Gaussian measurements~\cite{Lupo2020PRL_QuLinearLimits, Wang2025:GaussMeasurements}, such as those based on linear optics combined with photon counting.

\begin{figure}
    \centering
    \includegraphics[width=.25\linewidth]{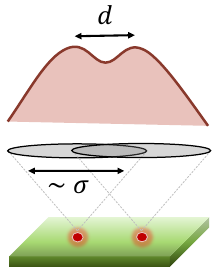}
    \caption{Resolving two emitters. PSF width $\sigma$ and separation distance $d$ with $d\ll\sigma$ in the subdiffraction regime. }
    \label{fig:twoemitters}
\end{figure}

To overcome Rayleigh's resolution limit (see Fig.~\ref{fig:twoemitters}), one can transition to a structured mode basis and count photons in that basis. Consider the expansion of the mutual coherence in the basis $\Psi = \{\psi_k\}$:
\begin{equation}\label{eq:app-gamma-basis}
    \Gamma(u,u') = \sum_{\ell,k} \Gamma_{\ell k}^{[\Psi]} \psi_\ell^*(u') \psi_k(u),
\end{equation}
where $\Gamma_{\ell k}^{[\Psi]} = \int \dd{u} \dd{u'} \psi_\ell(u) \psi_k^*(u') \Gamma(u,u')$. Defining the mode operators $\hat{\psi}_k = \int \dd{u} \psi_k(u) \hat{a}_u$, the passive imaging model in this basis becomes
\begin{equation}\label{eq:app-basis-rho}
    \rho(\Theta) \approx (1 - \varepsilon)\dyad{0} + \sum_{\ell,k} \Gamma_{\ell k}^{[\Psi]} \hat{\psi}_\ell^\dagger \dyad{0} \hat{\psi}_k.
\end{equation}
By tailoring the basis $\Psi$ to the imaging problem at hand, such as using Hermite-Gaussian (HG) modes adapted to a Gaussian PSF, one can optimally extract information at subdiffraction scales and overcome Rayleigh's curse (under the important assumption that the centroid is known). 

\subsection{Conventional imaging}

In conventional passive imaging, a spatially resolving detector (camera) collects photons and measures the intensity at each position $u$ on the collection plane. The probability to detect a single photon at location $u$ (conditioned on the occurrence of a detection event) is given by $p(u) = \Gamma(u,u)/\varepsilon$, which is proportional to the local intensity of the emissive scene on the collection plane. This yields a coarse-grained image of the scene because this approach fails to resolve fine (subdiffraction) features when emitters are smaller than the diffraction limit ($d\lesssim\sigma$).

To illustrate the resolution problem explicitly, consider the canonical problem of estimating the separation $d$ between two equally bright point sources with centroid fixed at the origin. From Eq.~\eqref{eq:app-gamma-uncorr}, the mutual coherence function reads
\begin{equation}\label{eq:2point-coherence}
    \Gamma(u,u') = \frac{\varepsilon}{2} \left[ \varphi(u - d/2)\varphi(u' - d/2) + \varphi(u + d/2)\varphi(u' + d/2) \right],
\end{equation}
which results in the intensity distribution
\begin{equation}
    p(u) = \frac{1}{2} \left[ |\varphi(u - d/2)|^2 + |\varphi(u + d/2)|^2 \right].
\end{equation}
We now expand $\varphi(u \pm d/2)$ in the subdiffraction regime $d \ll \sigma$:
\begin{equation}
    \varphi(u \pm d/2) \approx \varphi(u) \left(1 \mp \frac{ud}{4\sigma^2} - \frac{d^2}{16\sigma^2} + \frac{u^2d^2}{32\sigma^4}\right) + \order{\frac{d^3}{\sigma^3}}.
\end{equation}
Substituting into $p(u)$ and neglecting higher-order terms yields
\begin{equation}
    p(u) \approx  |\varphi(u)|^2 \left(1 - \frac{d^2}{8\sigma^2} + \frac{u^2d^2}{8\sigma^4}\right).
\end{equation}
The classical Fisher information (per photon) for estimating $d$ (in units of $\sigma$) from this distribution is then 
\begin{equation}
    \frac{F_{\rm direct}(d/\sigma)}{\varepsilon} = \int\dd{u} \frac{(\partial_d p(u))^2}{p(u)} \approx \frac{d^2}{8\sigma^2},
\end{equation}
which tends to zero in the limit $d \to 0$. This behavior captures the essence of Rayleigh's curse: Estimation precision vanishes for closely spaced incoherent sources under direct detection. In the next section, we demonstrate how this limitation can be overcome using structured measurements that are mode-resolved rather than pixel-resolved.

\subsection{Dispelling Rayleigh's curse}
\label{app:dispel-ray}

The resolution limit encountered in conventional imaging can be overcome by measuring in a spatial mode basis adapted to the structure of the imaging problem, as first discovered in the seminal work of Tsang \emph{et al.}~\cite{Tsang2016Superresolution}. This approach, termed spatial-mode demultiplexing (SPADE), circumvents Rayleigh's curse and enables subdiffraction imaging below Rayleigh's limit (with certain caveats, such as knowing the centroid~\cite{Tsang2016Superresolution}). We outline a simple calculation illustrating this result. See Ref.~\cite{Tsang2019Starlight} for a detailed review.

Given a Gaussian PSF, a natural choice for the mode basis is the HG mode basis,
\begin{equation}
    \psi_k(u) = \frac{1}{(2\pi\sigma^2)^{1/4}}\frac{1}{\sqrt{2^k k!}} H_k\left(\frac{u}{\sqrt{2}\sigma}\right) e^{-u^2 / (4\sigma^2)},
\end{equation}
where $H_k(x)$ are the physicist's Hermite polynomials with $H_0(x) = 1$, $H_1(x) = 2x$, $H_2(x) = 4x^2 - 2$, etc.

Consider again the two point-source separation problem in the subdiffraction regime $d \ll \sigma$. The mutual coherence function can be expanded in any mode basis as in Eq.~\eqref{eq:app-gamma-basis}. Since the PSF $\varphi(u \pm d/2)$ is a displaced Gaussian, we may write it in the HG basis using the well-known formula
\begin{equation}
    \varphi(u \pm d/2) = \sum_k c_k\left(\mp \frac{d}{2}\right) \psi_k(u),
\end{equation}
with
\begin{equation}\label{eq:app-ck}
    c_k\left(y\right) = \frac{1}{\sqrt{k!}} \left(\frac{y}{2\sigma}\right)^k e^{-y^2/8\sigma^2}.
\end{equation}
From here, we compute the HG-basis expansion coefficients [see Eq.~\eqref{eq:2point-coherence}] to second order in $d/\sigma$. Generally, we have 
\begin{equation}
    \Gamma^{[\Psi]}_{\ell k}=\frac{\varepsilon}{2}\Big(c_k(d/2)c_\ell(d/2)+c_k(-d/2)c_\ell(-d/2)\Big).
\end{equation}
Using Eq.~\eqref{eq:app-ck}, one can show that $\Gamma_{\ell k}^{[\Psi]}=\tfrac{\varepsilon}{\sqrt{k!\,l!}}e^{-d^{2}/(16\sigma^{2})}\left(\frac{d}{4\sigma}\right)^{k+l}$ for $k+\ell=$ even and $\Gamma_{\ell k}=0$ for $k+\ell=$ odd. Hence, the mutual coherence decouples into even-parity and odd-parity subspaces, due to inversion symmetry of the problem. The first non-negligible coefficients to order $d^2/\sigma^2$ are
\begin{equation}
    \Gamma_{00}^{[\Psi]} \approx \varepsilon \left( 1 - \frac{d^2}{16\sigma^2} \right), \quad
    \Gamma_{11}^{[\Psi]} \approx \frac{\varepsilon d^2}{16\sigma^2}, \quad
    \Gamma_{02}^{[\Psi]} \approx \frac{\varepsilon d^2}{16\sqrt{2} \sigma^2}, \quad
    \Gamma_{01}^{[\Psi]} = 0.
\end{equation}
We use these coefficients to write the output quantum state [see Eq.~\eqref{eq:app-basis-rho}] in the HG basis. If we restrict to mode-diagonal measurements (e.g., photon counting in the HG basis), the output state becomes
\begin{equation}
    \rho(\Theta) \approx (1 - \varepsilon)\dyad{0} + \varepsilon \left(1 - \frac{d^2}{16\sigma^2} \right) \hat{\psi}_0^\dagger \dyad{0} \hat{\psi}_0 + \frac{\varepsilon d^2}{16\sigma^2} \hat{\psi}_1^\dagger \dyad{0} \hat{\psi}_1 + \order{\frac{d^3}{\sigma^3}},
\end{equation}
where the off-diagonal component $\Gamma_{02}^{[\Psi]}$ is discarded due to HG-diagonal measurements.

The estimation strategy is now straightforward: Estimate the brightness, $\varepsilon$, from the total photon count, and estimate the separation, $d$, by counting photons in the antisymmetric $\psi_1$ mode. For brightness estimation, one may readily show that the classical Fisher information is $F(\varepsilon)\approx 1/\varepsilon$, which is quantum optimal given zero probe photons ($N_S=0$). For estimating the separation $d$, the Fisher information per detected photon in the mode $\psi_1$ is
\begin{equation}
    \frac{F_{\rm SPADE}(d/\sigma)}{\varepsilon} = \frac{\left(\partial_d p_1\right)^2}{p_1(1-p_1)} \approx \frac{1}{4}.
\end{equation}
This result is independent of $d$ to leading order and, furthermore, is quantum optimal~\cite{Tsang2016Superresolution}. Thus, SPADE dispels Rayleigh's curse and restores resolution in the subdiffraction regime. Note that the central idea is not limited to Gaussian PSFs. More generally, one can choose a mode basis adapted to the actual PSF of the imaging system; see Refs.~\cite{Rehacek2017pad,Kerviche2017pad} for further discussion on this. 

\paragraph*{\textbf{Digression.}} 
In the calculations above, we neglect contributions from the higher-order $\psi_2$ mode, as it carries negligible information about $\varepsilon$ or $d$. 
Although there is a small mixing between the $\psi_0$ and $\psi_2$ modes via the off-diagonal element $\Gamma_{02}^{[\Psi]}$, this mixing is perturbatively suppressed. 
To see this, we diagonalize the mutual coherence matrix in the even-parity $(\psi_0,\psi_2)$ subspace. 
The dominant eigenmode is $\psi_+\approx \psi_0 + d^2/(16\sqrt{2}\sigma^2)\psi_2$ with eigenvalue $\Lambda_+\approx \varepsilon (1-d^2/16\sigma^2)$. Whereas the orthogonal complement $\psi_-\approx -d^2/(16\sqrt{2}\sigma^2)\psi_0 + \psi_2$ (dominated by the $\psi_2$ mode) has a strongly suppressed eigenvalue, $\Lambda_-\propto d^4/\sigma^4$. Hence, the fundamental Gaussian mode $\psi_0$ overwhelmingly dominates the even-parity subspace.

\subsection{Beyond the ideal two-point-source problem}
\label{app:beyond-problems}

So far, we have focused on the canonical toy problem of estimating the separation between two equally bright, incoherent emitters. However, many other imaging problems, beyond the canonical two-point source separation problem, are of both practical and theoretical interest. We briefly highlight several directions, without going into much detail:  
\begin{enumerate}
    \item \textbf{Non-ideal or asymmetric sources.}
    Extensions to unequally bright emitters, unknown centroids, or other asymmetries~\cite{Rehacek2017MultiImaging,Fiderer2021GenExp_Imaging} lead to a genuine multiparameter estimation problem, often with incompatible parameters. Rayleigh’s curse typically re-emerges in these scenarios, although structured measurements still outperform direct detection in principle~\cite{Tsang2019Starlight}. Related problems include estimating separations in the presence of perturbative deformations (e.g., stresses or strains) in the emitter configuration~\cite{Sidhu2017SpatialDeform}.  
    
    \item \textbf{Localization.} Unsurprisingly, localizing emitters in all three spatial dimensions~\cite{Backlund2018localiz, Yu2018Imaging3D, Napoli2019SurfaceMetrology,Bisketzi2019QuLocaliz,Zhou2019zLocalization} admits a similar formalism as the 1D point-source separation problem, but requires estimating additional axial and lateral parameters.
    
    \item \textbf{Spatially correlated emitters.} Spatial emitter correlations [see end of Appendix \ref{app:emitter-probe}] can drastically affect precision limits. Larson \emph{et al.}~\cite{Larson:18correlation} showed that if correlations are strong, then it becomes challenging to estimate separation, in principle, due to inter-dependencies between the degree of correlation and the separation. Nonetheless, SPADE can perform better than conventional imaging techniques. 
    
    In a separate guise, macroscopic dissipative behavior, such as collective decay (a.k.a., superradiance)~\cite{Gross1982Superrad,Mok2024UniversalScaling}) or collective absorption~\cite{James2022SuperAbsorb}, in dense multi-emitter systems poses new challenges and questions. For instance, resolving superradiant versus subradiant channels may be challenging depending on mode structure of the quantum system being interrogated. 
    
    Beyond imaging, our methods may also prove useful for probing many-body phenomena with light~\cite{Nambiar2025:PhasesViaPhotonCorr}, where excitation signatures carrying information about two-point correlations can be inferred from the photon–photon correlations, in the regime of weak light–matter coupling that we consider herein.

    \item \textbf{Extended objects.} Going beyond point sources to reconstruct extended spatial structures involves estimating higher-order moments of an intensity distribution~\cite{Tsang2017NJP_ExtendedSource,Dutton2019PRA_ExtendedSource,Tsang2019PRA_Subdiffraction,Zhou2019PRA_ModernRayleigh}. In principle, this enables full characterization of object geometry and deformations.
    \item \textbf{Adaptive strategies.} In realistic multiparameter scenarios, imaging strategies may adapt dynamically as more information is gathered. Adaptive or Bayesian methods~\cite{Grace2020Prior,Lee2023Bayesian} and variational approaches~\cite{Feng2025:PhysLayerPCA} may prove fruitful here.
    \item \textbf{Hypothesis testing} Often, the goal is not parameter estimation but answering simpler, global questions such as whether there are one or two emitters. These tasks fall under hypothesis testing~\cite{Guha2018OnevsTwo,Huang2021Exoplanet,Grace2022Discrimination,Huang2023ExoSpectroscopy}, with possible extensions to multimode and multi-emitter/absorber scenes~\cite{Cox2024:quIllumination}. 
\end{enumerate} 
These problems illustrate the richness of quantum-limited imaging beyond the simple case of imaging two identical point emitters. Moreover, while most prior work has focused on passive strategies, active probe-based approaches, such as the structured twin-beam echo and structured Fock-state protocols that we develop in this work, offer a promising route to extend quantum advantages to more general and challenging settings.

\section{Gaussian Theory of Twin-Beam Echoes}
\label{app:echo}

In this appendix, we prove that the twin-beam echoes map the generalized imaging channel [Eq.~\eqref{eq:app-grand}] to two copies of the passive imaging channel [Eq.~\eqref{eq:app-passive-rho}] in the perturbative regime of weak emitter-probe interactions. The proof takes an elegant form in the Gaussian formalism by analyzing how covariance matrices transform under the relevant operations. Since this formalism and corresponding proofs deal purely with matrix manipulations, we use boldface $\bm{M}$ to denote matrices in this appendix and refer to their elements via indices, so that $\bm{M}_{ij}$ denotes the $(i,j)$ element of $\bm{M}$.

\subsection{Gaussian formalism in a nutshell}
\label{app:cov-bkgrd}

We work in the quadrature formalism, which provides a natural framework for describing Gaussian states. See Refs.~\cite{Weedbrook2012:GaussianRvw,Serafini2017:QCV} for more background on the Gaussian formalism.

Define the vector of quadrature operators for $M'$ modes as $\hat{\Vec{r}}= (\hat{q}_1,\hat{p}_1,\dots, \hat{q}_{M'},\hat{p}_{M'})$. Gaussian states are fully characterized by their first and second moments, namely the mean $\Vec{\mu}=\expval*{\hat{\Vec{r}}}$ and covariance $\bm{\Sigma}_{ij}=\expval*{\acomm*{\hat{\Vec{r}}_i-\vec{\mu}_i}{\hat{\Vec{r}}_j-\vec{\mu}_j}}$. Under a general Gaussian channel, $\Phi_{\mathscr{G}}$, the covariance matrix transforms as $\Phi_{\mathscr{G}}:\bm{\Sigma}\rightarrow \bm X\bm{\Sigma}\bm X^\top + \bm Y$ and the mean as $\Phi_{\mathscr{G}}:\vec{\mu}\rightarrow \bm X\vec{\mu}$, where $\bm X$ and $\bm Y$ are scaling and noise matrices, respectively~\cite{Serafini2017:QCV}. [We ignore coherent displacements which simply shift the mean.] For unitary Gaussian transformation (e.g., linear optical transformations and squeezers), $\bm Y=0$ and $\bm X=\bm S$, where $\bm S$ is a symplectic transformation. Importantly, Gaussian channels preserve Gaussianity; that is, they map Gaussian states to Gaussian states.

A (zero-mean) pure state satisfies $\bm{\Sigma}=\bm S \bm S^\top$ for some symplectic matrix $\bm S$. For instance, the vacuum state has $\bm{\Sigma}_{\rm vac}=\bm I$. A thermal state has covariance $\bm{\Sigma}_{\rm th}=\bm I + 2\bm Y_{\rm th}$ where $\bm Y_{\rm th}$ is the (positive semi-definite) thermal covariance that encodes classical correlations between the modes~\cite{Serafini2017:QCV}. In the single-photon approximation (e.g., low-temperature or high-frequency regime), a thermal state may be written explicitly as $\rho_{\rm th}\approx (1-\gamma)\rho_{\rm vac}+\sum_{i,j}\bm\gamma_{ij}\hat{a}_j^\dagger \rho_{\rm vac}\hat{a}_i$,
where $\gamma=\sum_i\bm\gamma_{ii}$ and $\bm\gamma_{ij}=\expval*{\hat{a}_i^\dagger\hat{a}_j}$ denotes the (Hermitian) mutual coherence. The mutual coherence $\bm\gamma$ may be related to the (real) quadrature covariance $\bm Y_{\rm th}$ by appropriate application of the unitary matrix $\bm{u}$ that transforms between quadrature and ladder operators~\cite{Serafini2017:QCV}, i.e. $(\hat{a},\hat{a}^\dagger)=\bm u (\hat{q},\hat{p})$ with $\bm u=\big(\begin{smallmatrix}
  1 & i\\
  1 & -i
\end{smallmatrix}\big)/\sqrt{2}$.

\subsection{Proof that twin-beam echoes imply passive imaging}
In passive imaging, an optical field in a (spatially correlated) thermal state is simply collected, processed, and measured. In this appendix, we show that implementing twin-beam echoes on signal-idler pairs initialized in vacuum [i.e., applying two-mode squeezing (and its inverse) prior to (and following) the signal-emitter interaction] maps the initial vacua to a multimode correlated thermal state, analogous to the passive imaging case. Since all operations involved in the echo sequence are Gaussian, we carry out the full analysis in the Gaussian covariance formalism.

Consider the following twin-beam echo sequence: 
\begin{itemize}
    \item[(1)] Generate $M'$ signal-idler twin beams using $M'$ unitary two-mode squeezing operations $\bigotimes_{i=1}^{M'} \hat{U}_{\rm tms}$. 
    \item[(2)] Interact signal modes (i.e., the probes) with the emitters, while the idlers are held in storage.
    \item[(3)] Apply the inverse two-mode squeezing operation $\bigotimes_{i=1}^{M'} \hat{U}_{\rm tms}^\dagger$. 
\end{itemize}
We now show that this sequence maps an initial vacuum state to a correlated thermal state.

In step (1), we generate $M'$ twin beams via $M'$ two-mode squeezing operations. The two-mode squeezing interactions are characterized by the ($4M' \times 4M'$) symplectic transformation $\bm{S}_{\rm tms}$, which acts on the covariance matrix of the $2M'$-mode vacuum state, $\bm{\Sigma}_{\rm vac} = \bm{I}_{4M'}$. [Note that each mode contributes two quadratures, so the total dimension is $4M'$.] We partition the covariance matrix into four blocks, with the upper-left and lower-right blocks corresponding to idler and signal modes, respectively, and off-diagonal blocks encoding signal-idler correlations. After step (1), the covariance matrix is
\begin{equation}
    \bm{\Sigma}_1 = \bm{S}_{\rm tms} \bm{\Sigma}_{\rm vac} \bm{S}_{\rm tms}^\top = \bm{S}_{\rm tms} \bm{S}_{\rm tms}^\top,
\end{equation}
and we write the symplectic transformation for pairwise (a.k.a., two-mode) squeezing in block form as
\begin{equation}
    \bm{S}_{\rm tms} = \begin{pmatrix}
        \bm{\alpha} & \bm{\beta} \\
        \bm{\beta} & \bm{\alpha}
    \end{pmatrix},
\end{equation}
where $\bm\alpha^\top\bm\beta - \bm\beta^\top\bm\alpha=0$ and $\bm\alpha^\top\bm\alpha-\bm\beta^\top\bm\beta=\bm I_{2M'}$. For general pairwise squeezing, we have $\bm{\alpha} = \bigoplus_{i=1}^{M'} \cosh(r_i) \bm{I}_2$ and $\bm{\beta} = \bigoplus_{i=1}^{M'} \sinh(r_i) \bm{R}(\phi_i)$, where $\bm{R}(\phi_i)=\big(\begin{smallmatrix}
  \cos\phi_i & \sin\phi_i\\
  \sin\phi_i & -\cos\phi_i
\end{smallmatrix}\big)$. This realization of $\bm{\alpha}$ and $\bm{\beta}$ is not needed in what follows but is nevertheless representative.

We now move to step (2), which involves emitter-probe interactions. Since this interaction is restricted to the signal modes (idler modes are held in storage), the interaction may be described by the scaling and noise matrices
\begin{equation}
\bm{X} = \bm{I}_{2M'} \oplus \bm{X}_S \qq{and} \bm{Y} = \bm{0}_{2M'} \oplus \bm{Y}_S.
\end{equation}
In the perturbative regime of weak interactions, we approximate
\begin{equation}
    \bm{X}_S \approx \bm{I}_{2M'} + \frac{\varsigma}{2} \bm{\eta} \qq{and} \bm{Y}_S = \bm{\eta},
\end{equation}
where $\varsigma = -1$ for absorption (pure-loss channel) and $\varsigma = +1$ for incoherent emission (quantum-limited amplifier channel), and $\bm{\eta}$ represents the (positive, semi-definite) perturbation. [If both absorption and emission are present, then we treat them perturbatively and include both via $\bm{X}_S \approx \bm{I}_{2M'} + (\bm{\eta}_{\downarrow} - \bm{\eta}_{\uparrow})/2$.] Applying this Gaussian channel to $\bm{\Sigma}_1$ yields the post-interaction covariance
\begin{align}
    \bm{\Sigma}_2 &= \bm{X} \bm{\Sigma}_1 \bm{X}^\top + \bm{Y} \\
    &\approx \bm{\Sigma}_1 + \frac{\varsigma}{2} \left[ (\bm{0} \oplus \bm{\eta}) \bm{\Sigma}_1 + \bm{\Sigma}_1 (\bm{0} \oplus \bm{\eta}) \right] + (\bm{0} \oplus \bm{\eta}) \\
    &= \bm{S}_{\rm tms} \bm{S}_{\rm tms}^\top + \frac{\varsigma}{2} \left[ (\bm{0} \oplus \bm{\eta}) \bm{S}_{\rm tms} \bm{S}_{\rm tms}^\top + \bm{S}_{\rm tms} \bm{S}_{\rm tms}^\top (\bm{0} \oplus \bm{\eta}) \right] + (\bm{0} \oplus \bm{\eta}),
\end{align}
which is valid to first order in $\bm{\eta}$.

In step (3), we apply the inverse two-mode squeezing operation to nearly unentangle the twin-beams. The inverse two-mode squeezing operation is represented by the symplectic matrix
\begin{equation}
    \bm{S}_{\rm tms}^{-1} = \begin{pmatrix}
        \bm{\alpha} & -\bm{\beta} \\
        -\bm{\beta} & \bm{\alpha}
    \end{pmatrix}.
\end{equation}
Applying this transformation to $\bm{\Sigma}_2$ yields the final covariance matrix 
\begin{align}
    \bm{\Sigma}_3 &= \bm{S}_{\rm tms}^{-1} \bm{\Sigma}_2 \bm{S}_{\rm tms}^{-\top} \\
    &\approx \bm{I}_{4M'} + \frac{\varsigma}{2} \left[ \bm{S}_{\rm tms}^{-1} (\bm{0} \oplus \bm{\eta}) \bm{S}_{\rm tms} + \bm{S}_{\rm tms}^\top (\bm{0} \oplus \bm{\eta}) \bm{S}_{\rm tms}^{-\top} \right] + \bm{S}_{\rm tms}^{-1} (\bm{0} \oplus \bm{\eta}) \bm{S}_{\rm tms}^{-\top}\\   
    &= \bm I_{4M'} + 2\,\bm{\delta Y}_{\rm th}
\end{align}
where we define the thermal perturbation,
\begin{equation}
    \bm{\delta Y}_{\rm th} = \frac{\varsigma}{4} \left[ \bm{S}_{\rm tms}^{-1} (\bm{0} \oplus \bm{\eta}) \bm{S}_{\rm tms} + \bm{S}_{\rm tms}^\top (\bm{0} \oplus \bm{\eta}) \bm{S}_{\rm tms}^{-\top} \right] + \frac{1}{2} \bm{S}_{\rm tms}^{-1} (\bm{0} \oplus \bm{\eta}) \bm{S}_{\rm tms}^{-\top}.\label{eq:dYth}
\end{equation}
We conclude that the output state is approximately a $2M'$-mode correlated thermal state to first order in $\bm\eta$.

\subsection{Composition of signal and idler sectors}

We now analyze the structure of the final output covariance $\bm{\Sigma}_3 = \bm{I}_{4M'} + 2\bm{\delta Y}_{\rm th}$ obtained after the twin-beam echo sequence. The quantity $\bm{\delta Y}_{\rm th}$ [see Eq.~\eqref{eq:dYth}] captures the thermal perturbation which encodes the absorption and fluorescence features, as well as amplification pre-factors from the twin-beam echoes. 

For clarity, we evaluate $\bm{\delta Y}_{\rm th}$ term by term. First,
\begin{align}
    \bm{S}_{\rm tms}^{-1}(\bm{0}\oplus\bm{\eta})\bm{S}_{\rm tms} 
    &= 
    \begin{pmatrix}
        \bm{\alpha} & -\bm{\beta} \\
        -\bm{\beta} & \bm{\alpha}
    \end{pmatrix}
    \begin{pmatrix}
        \bm{0} & \bm{0} \\
        \bm{0} & \bm{\eta}
    \end{pmatrix}
    \begin{pmatrix}
        \bm{\alpha} & \bm{\beta} \\
        \bm{\beta} & \bm{\alpha}
    \end{pmatrix} \\
    &=
    \begin{pmatrix}
        -\bm{\beta}\bm\eta\bm\beta &
        -\bm{\beta}\bm\eta\bm\alpha \\
        \bm{\alpha}\bm\eta\bm\beta
        & \bm{\alpha}\bm\eta\bm\alpha
    \end{pmatrix}.
\end{align}
Using the identity
\begin{equation}
    \bm{S}_{\rm tms}^{-1}(\bm{0}\oplus\bm{\eta})\bm{S}_{\rm tms} = 
    \left[ \bm{S}_{\rm tms}^\top (\bm{0}\oplus\bm{\eta}) \bm{S}_{\rm tms}^{-\top} \right]^\top,
\end{equation}
the symmetric contribution in Eq.~\eqref{eq:dYth} simplifies to
\begin{equation}
    \frac{\varsigma}{4} \left( \bm{S}_{\rm tms}^{-1} (\bm{0} \oplus \bm{\eta}) \bm{S}_{\rm tms} + \bm{S}_{\rm tms}^\top (\bm{0} \oplus \bm{\eta}) \bm{S}_{\rm tms}^{-\top} \right) 
    =
    \frac{\varsigma}{2}
    \begin{pmatrix}
        -\bm{\beta}\bm{\eta}\bm{\beta} & \bm{0} \\
        \bm{0} & \bm{\alpha}\bm{\eta}\bm{\alpha}
    \end{pmatrix}.
    \label{eq:dV_symmetric}
\end{equation}

The last term in Eq.~\eqref{eq:dYth} evaluates straightforwardly to
\begin{equation}
    \frac{1}{2} \bm{S}_{\rm tms}^{-1} (\bm{0} \oplus \bm{\eta}) \bm{S}_{\rm tms}^{-\top} = \frac{1}{2}
    \begin{pmatrix}
        \bm{\beta}\bm\eta\bm\beta &
        -\bm{\beta}\bm\eta\bm\alpha \\
        -\bm{\alpha}\bm\eta\bm\beta
        & \bm{\alpha}\bm\eta\bm\alpha
    \end{pmatrix}.
    \label{eq:dV_remainder}
\end{equation}
Combining Eqs.~\eqref{eq:dV_symmetric} and~\eqref{eq:dV_remainder}, we obtain the final expression for the thermal perturbation:
\begin{equation}
    \bm{\delta Y}_{\rm th} = \frac{1}{2}
    \begin{pmatrix}
         (1-\varsigma)\bm{\beta}\bm{\eta}\bm{\beta} & -\bm{\beta}\bm{\eta}\bm{\alpha} \\
          -\bm{\alpha}\bm{\eta}\bm{\beta} &  (1+\varsigma)\bm{\alpha}\bm{\eta}\bm{\alpha}
    \end{pmatrix}.
\end{equation}
The upper-left and lower-right blocks of $\bm{\delta Y}_{\rm th}$ describe thermal excitations within the idler and signal sectors, respectively, while the off-diagonal blocks encode thermal correlations between the two sectors.

For $\varsigma = -1$ (absorption), only the idler sector acquires thermal excitations, while the signal remains in vacuum. Conversely, for $\varsigma = +1$ (incoherent emission), the signal sector becomes thermally excited, while the idler remains in vacuum. Thus we have shown that twin-beam echoes map weak emitter-probe interactions onto thermal excitations localized within either the signal or idler sectors, depending on the physical nature of the interaction.

The off-diagonal terms in $\bm{\delta Y}_{\rm th}$  encode residual phase-sensitive signal-idler correlations, but are irrelevant if we perform excitation measurements on the signal and idler sectors independently. That is, we consider measurement operators of the form $\{ \Pi_{N_S} \otimes \Pi_{N_I} \}$, where $\Pi_{N_S}$ is a projection onto the total $N_S$-excitation subspace of the signal modes, and similarly for $\Pi_{N_I}$. These measurements effectively erase inter-sector (signal-idler) coherences, while preserving intra-sector (signal-signal and idler-idler) coherences and the overall thermal statistics. Formally, the measurement-averaged state is thus 
\begin{equation}
    \bar{\rho}_{SI}^{\,\rm{echo}} = \sum_{N_S, N_I} 
    (\Pi_{N_S} \otimes \Pi_{N_I})\, 
    \rho_{SI}^{\rm{echo}}\,
    (\Pi_{N_S} \otimes \Pi_{N_I}) = \rho_S^{\rm{echo}} \otimes \rho_I^{\rm{echo}},
\end{equation}
where $\rho_S^{\rm{echo}}$ and $\rho_I^{\rm{echo}}$ are thermal states supported on the signal and idler sectors, respectively, with corresponding covariances
\begin{equation}
    \bm{\Sigma}_S^{\rm echo} = \bm I_{2M'} + (1+\varsigma)\bm{\alpha}\bm{\eta}\bm{\alpha} \qq{and}
\bm{\Sigma}_I^{\rm echo} = \bm I_{2M'} + (1-\varsigma)\bm{\beta}\bm{\eta}\bm{\beta}.
\end{equation}
If both absorption and emission are present, then we simply have $\bm{\Sigma}_S^{\rm echo} = \bm I_{2M'} + 2\bm{\alpha}\bm{\eta}_{\downarrow}\bm{\alpha}$ and $\bm{\Sigma}_I^{\rm echo} = \bm I_{2M'} + 2\bm{\beta}\bm{\eta}_{\uparrow}\bm{\beta}$, with $\bm{\eta}_{\uparrow/\downarrow}$ representing absorption/emission profiles in the quadrature-covariance formalism.

\section{Ultimate Sensitivity Limits}
\label{app:sensitivity-limits}

In this appendix, we provide quantitative insight into the ultimate sensitivity limits of various imaging and sensing tasks that we consider in this work. These limits are determined by the Quantum Fisher Information (QFI), which sets the fundamental bound on parameter-estimation precision~\cite{Paris2009QFI,Sidhu2020QFIrvw}. We first review single-parameter bounds for simple (single-mode) estimation problems where QFI results are well established, such as absorption~\cite{Monras2007LossEst,Adesso2009LossEst}, gain estimation~\cite{Aspachs2010:UnruhEst}, and random-displacement estimation~\cite{Gorecki2022SpreadChannel,Shi2023DMLimits}. We then turn to the multimode scenario of subdiffraction imaging with quantum probes, where we consider pure absorption and pure incoherent-emission channels separately. For a classical baseline comparison, we also provide coherent-state sensitivity bounds. Together, these examples provide strong evidence that our universal quantum imaging module not only offers enhanced performance over classical strategies in multiparameter settings but also operates at or near the fundamental sensitivity limits imposed by quantum mechanics. The sensitivity bounds, including new bounds derived in this work, are collected in Table~\ref{tab:qfi_loss_gain_imaging} for convenience.

\subsection{Known limits of channel estimation}
\label{app:channel-estimation}

Here we summarize known sensitivity bounds for three canonical bosonic channels: the pure-loss channel~\cite{Monras2007LossEst,Adesso2009LossEst}, the quantum-limited amplifier channel~\cite{Aspachs2010:UnruhEst}, the stochastic displacement (or additive Gaussian noise) channel~\cite{Gorecki2022SpreadChannel,Shi2023DMLimits}, and more broadly, phase-covariant bosonic channels~\cite{Nair2023:CovChSensing}. These single-mode results are well established in quantum metrology and serve as useful references for the more complex multimode imaging problems. 

To bound the QFI for estimating the parameter $\theta$ encoded through the (non-unitary) quantum channel $\mathcal{E}_\theta$, we invoke the concept of a unitary dilation (or unitary extension). In particular, any valid quantum channel admits a unitary dilation,  $\mathcal{E}_\theta(\rho_S)=\Tr_E\{\hat{U}_{SE}(\theta)\rho_S\otimes\tau_E\hat{U}_{SE}(\theta)\}$, where $\tau_E$ is a fixed environment state that may be taken pure and $\hat{U}_{SE}(\theta)$ is a unitary acting jointly on the system and environment that  encodes $\theta$. For this dilation, the QFI for estimating $\theta$ is then upper-bounded by the variance of the generator, $\hat{H}\coloneqq i (\partial_\theta \hat{U}_{SE})\hat{U}_{SE}^\dagger$, evaluated on the joint input state, $\rho_S\otimes\tau_E$. That is, $F_Q(\theta)\leq 4 {\rm Var}(\hat{H})$~\cite{Giovannetti2006:quMetrology,Fujiwara2008:FibreBundle,Escher2011:GenFramework}. 

\paragraph*{\textbf{Loss estimation.}}
The task is to estimate the transmittance $\eta$ of a bosonic pure-loss channel $\mathcal{L}_\eta$~\cite{Monras2007LossEst,Adesso2009LossEst}. In the Heisenberg picture, the channel $\mathcal{L}_\eta$ transforms the annihilation operator $\hat{a}$ via $\hat{a} \to \sqrt{\eta} \hat{a} + \sqrt{1-\eta} \hat{e}$, where $\hat{e}$ is the annihilation operator of an environment mode in the vacuum state, and $0\leq \eta \leq 1 $. We parametrize the transmittance as $\eta = e^{-\gamma_{\rm loss}}$, where $\gamma_{\rm loss}$ represents the loss rate (normalized by the inverse time of the interaction), analogous to the absorption rate $\gamma_{\uparrow}$ of this work.

Given an input state $\rho_S$, the Kraus (or operator-sum) representation of the pure-loss channel is
\begin{equation}
\mathcal{L}_\eta(\rho_S) = \sum_{n=0}^\infty \hat{K}_n \rho_S \hat{K}_n^\dagger,
\label{eq:kraus_loss}
\end{equation}
with Kraus operators~\cite{Ivan2011:OpSum,Gagatsos2017:OpSum}
\begin{equation}
\hat{K}_n = \left(\sqrt{1-\eta} \right)^n\eta^{\hat{a}^\dagger\hat{a}/2} \frac{\hat{a}^n}{\sqrt{n!}},
\label{eq:kraus-ops-loss}
\end{equation}
This representation arises from its purification to a two-mode beamsplitter interaction, $\hat{U}_\theta=\exp[\theta(\hat{a}^\dagger\hat{e}-\hat{a}\hat{e}^\dagger)]$, acting on the signal state $\rho_S$ and an environment mode in vacuum. The beamsplitter angle $\theta$ and transmittance $\eta$ are related via $\eta=\cos^2\theta$. Observe that $\hat{H}_{\rm loss}=i(\hat{a}^\dagger\hat{e}-\hat{a}\hat{e}^\dagger)$ represents an effective Hamiltonian of the interaction. In this manner, $\mathcal{L}_\eta(\rho_S)=\Tr_E\{\hat{U}_\theta\rho_S\otimes\dyad{0}_E\hat{U}_\theta^\dagger\}$, where $\dyad{0}_E$ denotes the environmental vacuum. Note that the perturbative regime analyzed throughout this work corresponds to keeping only the $n=0,1$ terms of the Kraus representation under the assumption that $(1-\eta)\expval*{\hat{a}^\dagger\hat{a}}\ll 1$.

From here, we derive the sensitivity bound set by the QFI. Suppose that we had access to the environment modes and that $\rho_S$ is a pure state (by convexity of the QFI, pure states suffice). Estimation then reduces to standard unitary parameter estimation for $\theta$. In this case, the QFI is bounded by the generator variance~\cite{Giovannetti2006:quMetrology} $F_Q(\theta)\leq 4{\rm Var}({\hat{H}_{\rm loss}})\leq 4N_S$, where $N_S=\expval*{\hat{a}^\dagger\hat{a}}$ is the mean photon number of the (pure) probe state $\rho_S$. Reparametrizing via $\eta=\cos^2\theta$ and invoking the reparametrization rule for the Fisher information [i.e., $F(\phi)=(\partial\varphi/\partial\phi)^2F(\varphi)$], we find $F_Q(\eta)=N_S/[\eta(1-\eta)]$, in agreement with Ref.~\cite{Adesso2009LossEst}. Finally, using $\eta=e^{-\gamma_{\rm loss}}$, it follows that
\begin{equation}\label{eq:qfi-loss}
    F_Q(\gamma_{\rm loss})=\frac{N_S e^{-\gamma_{\rm loss}}}{1-e^{-\gamma_{\rm loss}}},
\end{equation}
consistent with Table~\ref{tab:qfi_loss_gain_imaging} in the weak-loss limit ($\gamma_{\rm loss}\ll 1$). This bound is achieved by Fock states~\cite{Adesso2009LossEst} and two-mode squeezed-vacuum states~\cite{Nair2018:LossEst} for all values of $\eta$, as well as single-mode squeezed-vacuum states in the weak-absorption limit~\cite{Monras2007LossEst}. For two-mode squeezed-vacuum states, an actual receiver design is known~\cite{Gong2023:TransmitSens}.

\paragraph*{\textbf{Amplification estimation.}}
The task is to estimate the gain $G$ of a quantum-limited, phase-insensitive amplifier channel $\mathcal{A}_G$~\cite{Aspachs2010:UnruhEst}. In the Heisenberg picture, the channel $\mathcal{A}_G $ transforms the annihilation operator $ \hat{a} $ via $ \hat{a} \to \sqrt{G} \hat{a} + \sqrt{G - 1} \hat{e}^\dagger $, where $ \hat{e} $ is the annihilation operator of an environment mode in the vacuum state, and $ G \geq 1 $. We parametrize the gain as $ G = e^{\gamma_{\rm amp}}$, where $ \gamma_{\rm amp} $ represents the (normalized) amplification rate, analogous to the incoherent-emission rate $\gamma_{\downarrow}$ of this work.

Given an input state $\rho_S$, the Kraus (or operator-sum) representation of the quantum-limited amplifier channel is
\begin{equation}
\mathcal{A}_G(\rho_S) = \sum_{n=0}^\infty \hat{K}_n \rho_S \hat{K}_n^\dagger,
\label{eq:kraus_amp}
\end{equation}
with Kraus operators~\cite{Ivan2011:OpSum,Gagatsos2017:OpSum}
\begin{equation}
\hat{K}_n = \sqrt{\frac{1}{G}} \left( \sqrt{\frac{G - 1}{G}} \right)^n \frac{(\hat{a}^\dagger)^n}{\sqrt{n!}}G^{-\hat{a}^\dagger \hat{a}/2}.
\label{eq:kraus-ops-gain}
\end{equation}
This representation arises from its purification to a two-mode squeezing interaction, $\hat{U}_r=\exp[r(\hat{a}^\dagger\hat{e}^\dagger-\hat{a}\hat{e})]$, acting on the signal state $\rho_S$ and an environment mode in vacuum. The squeezing strength $r$ and gain $G$ are related via $G=\cosh^2r$. Observe that $\hat{H}_{\rm amp}=i(\hat{a}^\dagger\hat{e}^\dagger-\hat{a}\hat{e})$ represents an effective Hamiltonian of the interaction. In this manner, $\mathcal{A}_G(\rho_S)=\Tr_E\{\hat{U}_r\rho_S\otimes\dyad{0}_E\hat{U}_r^\dagger\}$, where $\dyad{0}_E$ denotes the environmental vacuum. Note that the perturbative regime analyzed throughout this work corresponds to keeping only the $n=0,1$ terms of the Kraus representation under the assumption that $(G-1)\expval*{\hat{a}^\dagger\hat{a}+1}\ll 1$.

From here, we derive the sensitivity bound set by the QFI. Suppose we had access to the environment modes and that $\rho_S$ is a pure state. Estimation then reduces to standard unitary parameter estimation for $\theta$. For which the QFI is bounded by the generator variance~\cite{Giovannetti2006:quMetrology} $F_Q(r)\leq 4{\rm Var}({\hat{H}_{\rm amp}})\leq 4(N_S+1)$, where $N_S=\expval*{\hat{a}^\dagger\hat{a}}$ is the mean photon number of the probe state $\rho_S$. Reparametrizing via $G=\cosh^2r$ and invoking the reparametrization rule for the Fisher information, we find that $F_Q(G)=(1+N_S)/G(G-1)$, in agreement with Ref.~\cite{Aspachs2010:UnruhEst}. Finally, using $G=e^{\gamma_{\rm amp}}$, it follows that
\begin{equation}\label{eq:qfi-gain}
    F_Q(\gamma_{\rm amp})=\frac{(1+N_S)e^{\gamma_{\rm amp}}}{e^{\gamma_{\rm amp}}-1},
\end{equation}
consistent with Table~\ref{tab:qfi_loss_gain_imaging} in the weak-amplification limit ($\gamma_{\rm amp}\ll 1$). This bound is achieved by Fock-state probes~\cite{Aspachs2010:UnruhEst} and, by extension, two-mode squeezed-vacuum states, as well as single-mode squeezed-vacuum states when $\gamma_{\rm amp}\ll 1$.

\paragraph*{\textbf{Stochastic displacement estimation.}}
We compute a perturbative QFI bound for the (single-mode) additive Gaussian noise channel $\Phi_{\gamma_{\rm agn}}$ in the limit of small additive noise, $\gamma_{\rm agn} \ll 1$. In the Heisenberg picture, the channel acts as ${\Phi_{\gamma_{\rm agn}}:\hat{a} \to \hat{a} + \beta}$, where $\beta \sim \mathcal{CN}(0, \gamma_{\rm agn})$ and $\gamma_{\rm agn}$ denotes the mean number of photons added to the state.

The additive Gaussian noise channel can be decomposed into a concatenation of a pure-loss channel followed by a quantum-limited amplifier~\cite{Weedbrook2012:GaussianRvw}, $\Phi_{\gamma_{\rm agn}} = \mathcal{A}_G \circ \mathcal{L}_\eta$, at the special point $G = 1/\eta \eqqcolon 1 + \gamma_{\rm agn}$. This construction formally admits a unitary dilation
\begin{equation}
    \Phi_{\gamma_{\rm agn}}(\rho_S) = \Tr_{E_1E_2} \left\{ \hat{U}_{\rm agn} \left( \rho_S \otimes \dyad{0}_{E_1} \otimes \dyad{0}_{E_2} \right) \hat{U}_{\rm agn}^\dagger \right\},
\end{equation}
where $\dyad{0}_{E_k}$ denotes the vacuum on the environmental subsystems and $\hat{U}_{\rm agn}\coloneqq \hat{U}_{SE_2}(r) \hat{U}_{SE_1}(\theta)$ denotes the joint unitary that implements a beam splitter (pure-loss dilation) followed by two-mode squeezing (quantum-limited amplifier dilation). The parameters $\theta$ and $r$ are defined via $\eta = \cos^2\theta$ and $G = \cosh^2 r$. As before, the beamsplitter and two-mode squeezing unitaries take the forms
$\hat{U}_{SE_1}(\theta) = \exp[\theta (\hat{a}^\dagger \hat{e}_1 - \hat{a} \hat{e}_1^\dagger)]$ and $\hat{U}_{SE_2}(r) = \exp[r (\hat{a}^\dagger \hat{e}_2^\dagger - \hat{a} \hat{e}_2)]$.

Given $\gamma_{\rm agn} \ll 1$, we expand $\tan^2 \theta = \gamma_{\rm agn}$ and $\sinh^2 r = \gamma_{\rm agn}$ via $\theta \approx \sqrt{\gamma_{\rm agn}} + \order*{\gamma_{\rm agn}^{3/2}}$ and ${r \approx \sqrt{\gamma_{\rm agn}} + \order*{\gamma_{\rm agn}^{3/2}}}$. Using the Baker–Campbell–Hausdorff formula ($e^{\hat{X}}e^{\hat{Y}}=e^{\hat{Z}}$ where $\hat{Z}=\hat{X}+\hat{Y}+\comm*{\hat{X}}{\hat{Y}}/2+\dots$) for the joint unitary $\hat{U}_{\rm agn}$, and retaining only first-order terms, we find
\begin{equation}
    \hat{U}_{\rm agn} \approx \exp\left[ \sqrt{\gamma_{\rm agn}} \left( \hat{a}^\dagger \hat{e}_1 - \hat{a} \hat{e}_1^\dagger + \hat{a}^\dagger \hat{e}_2^\dagger - \hat{a} \hat{e}_2 \right) + \order{\gamma_{\rm agn}} \right].
\end{equation}
Define the generator $\hat{H}_{\rm agn} = i (\partial_\gamma \hat{U}_{\rm agn}) \hat{U}_{\rm agn}^\dagger$, such that
\begin{equation}
    \hat{H}_{\rm agn} \approx \frac{i}{2\sqrt{\gamma_{\rm agn}}} \left( \hat{a}^\dagger \hat{e}_1 - \hat{a} \hat{e}_1^\dagger + \hat{a}^\dagger \hat{e}_2^\dagger - \hat{a} \hat{e}_2 \right).
\end{equation}
It follows straightforwardly that the QFI with access to the full-system $SE_1E_2$ is approximately $F_Q(\gamma_{\rm agn}; SE_1E_2) \approx 4\,\mathrm{Var}(\hat{H}_{\rm agn})$. Substituting the state $\rho_S \otimes \dyad{0}_{E_1} \otimes \dyad{0}_{E_2}$ then yields $\mathrm{Var}(\hat{H}_{\rm agn}) = \frac{1 + 2N_S}{4\gamma_{\rm agn}}$, where $N_S$ is the mean photon number of the probe $\rho_S$. Therefore, given only access to the subsystem $S$, the QFI is bounded by
\begin{equation}\label{eq:qfi-agn}
    F_Q(\gamma_{\rm agn}) \lesssim \frac{1 + 2N_S}{\gamma_{\rm agn}},
\end{equation}
consistent with the expression quoted in Table~\ref{tab:qfi_loss_gain_imaging}.

In the perturbative regime, Fock states~\cite{Wolf2019:MotionalFock}, single-mode squeezed vacuum~\cite{Gorecki2022SpreadChannel,Tsang2023NoiseSpectr}, and ancilla-assisted two-mode squeezed vacuum~\cite{Shi2023DMLimits} saturate the QFI (see also Ref.~\cite{Nair2023:CovChSensing}). Though, in the presence of spurious noise, such as heating or excitation loss, single-mode squeezed vacuum becomes strictly suboptimal~\cite{Shi2023DMLimits,Gardner2025StochWaveform}, whereas Fock states or two-mode squeezed vacuum (with perfect idler storage) remain optimal~\cite{Shi2023DMLimits,Gardner2025StochWaveform}.

\subsection{QFI limits for subdiffraction imaging}
\label{app:subdiff-qfi}

We now turn to ultimate sensitivity bounds for estimating the separation between two point-like objects in the subdiffraction regime ($d\ll\sigma$) with non-trivial quantum probes. In this regime, this problem reduces to a single-mode channel-estimation task, with the relevant loss (or gain) parameters encoded predominantly in the antisymmetric (e.g., the $n=1$ HG) mode~\cite{Tsang2016Superresolution,Lupo2016PRL_SubwaveImaging}. Thus, the QFI scalings follow straightforwardly from the canonical single-mode channel-estimation bounds summarized in Appendix~\ref{app:channel-estimation}. 

We consider two toy problems: pure absorption imaging (no incoherent emission) and pure fluorescence imaging (no absorption). These establish benchmarks for the more general imaging problem containing both fluorescence and absorption, and, moreover, help to clarify the performance of our quantum imaging module relative to fundamental sensitivity limits.

\paragraph*{\textbf{Subdiffraction absorption imaging.}}  
Consider two identical absorbers, each characterized by the absorption rate $\gamma_\uparrow$ and separated by a distance $d$, that we interrogate with quantum optical probes; we assume the centroid $x_1+x_2=0$ is known. The probes interact with the absorbers (assumed to be in their ground states), and the transmitted light then propagates to the collection plane. Hence, we are dealing with a purely absorption imaging problem with only two absorption points.

In the perturbative regime, the signal modes evolve under the general absorption channel,
\begin{equation}
    \rho_S(\Theta) \approx \rho_S +  \int\dd{u}\dd{u'}\Gamma_{\uparrow}(u,u')\left(\hat{a}_u\rho_S\hat{a}_{u'}^\dagger-\frac{1}{2}\acomm{\hat{a}_{u'}^\dagger\hat{a}_u}{\rho}\right),
\end{equation}
with absorption kernel
\begin{equation}\label{eq:abs-kernel-2point}
    \Gamma_{\uparrow}(u,u')=\gamma_\uparrow \left[ \varphi(u - d/2)\varphi(u' - d/2) + \varphi(u + d/2)\varphi(u' + d/2) \right].
\end{equation}
The absorption kernel has the mathematical form of the mutual coherence for the two-point source fluorescence imaging problem (see Appendix~\ref{app:imaging}). Therefore, the same symmetric/antisymmetric mode decomposition that diagonalizes the mutual coherence in the passive imaging setup (cf.~Refs.~\cite{Tsang2016Superresolution,Lupo2016PRL_SubwaveImaging}) also diagonalizes the absorption kernel. The multimode channel then decouples into two independent bosonic loss channels that can be probed independently for optimal estimation~\cite{Nair2018:LossEst}. In the subdiffraction regime ($d\ll \sigma$), the absorption in the antisymmetric mode, quantified by $\Gamma_{\uparrow,-}$, carries all the information about the emitter separation $d$. For a Gaussian PSF, the value of $\Gamma_{\uparrow,-}$ coincides with the HG1 coefficient, $\Gamma_{\uparrow,-}=\Gamma_{\uparrow,11}=\gamma_\uparrow d^2/(8\sigma^2)$, which derives from the same decomposition presented in Appendix~\ref{app:dispel-ray}. 

Finally then, using the general QFI for a pure-loss channel with transmittance $\eta_{-}\approx 1-\Gamma_{\uparrow,-}$~\cite{Monras2007LossEst,Adesso2009LossEst} [see also Eq.~\eqref{eq:qfi-loss}], we reckon the QFI for estimating the separation between the two point absorbers: 
\begin{equation}
F_Q\bigl(d/\sigma\bigr) \approx \frac{\gamma_\uparrow}{2} N_S,
\end{equation}
where $N_S$ is the total probe photon number and $\sigma$ is the PSF width. 

Our twin-beam echo protocol in the HG mode basis (as well as structured Fock states), followed by photon-counting measurements on the idler modes, attains this bound in the subdiffraction regime, supporting the optimality of our quantum imaging module. The separation $d$ can be measured by counting photons associated with the antisymmetric (a.k.a., HG1) mode, exactly as in passive fluorescence imaging. If $\gamma_\uparrow$ must also be estimated, then both the symmetric mode (essentially the HG0 mode, $\psi_0$) and the antisymmetric mode (the HG1 mode, $\psi_1$) should be addressed, thus enabling simultaneous estimation. If both $\gamma_\uparrow$ and $d$ are unknown, a fixed signal-photon budget must be divided between the symmetric and antisymmetric modes to estimate them jointly. Consequently, the respective single-parameter QFI optima cannot be attained simultaneously at fixed $N_S$.

We note that single-mode squeezed-vacuum probes are also optimal for weak absorption imaging~\cite{Monras2007LossEst}, however its use is severely limited when both absorption and fluorescence (or a small amount of Gaussian noise) are present, as described in Section~\ref{sec:single-sqz} of the main text.

\paragraph*{\textbf{Subdiffraction fluorescence imaging.}}
We cook up a similar toy example as the subdiffraction imaging problem above but now restricting to incoherent emission only. Consider two identically fluorescing emitters separated by a distance $d$ and of known centroid $x_1+x_2=0$. This toy example serves to illustrate the ultimate performance limit of fluorescence imaging by allowing us to simplify the setup and effectively map to amplifier-channel estimation. 

In the perturbative regime, the signal modes evolve under the general amplifier channel,
\begin{equation}
    \rho_S(\Theta) \approx \rho_S + \int\dd{u}\dd{u'} \Gamma_\downarrow(u,u') \Big( \hat{a}_{u'}^\dagger \rho \hat{a}_u - \frac{1}{2} \acomm{\hat{a}_u \hat{a}_{u'}^\dagger}{\rho} \Big),
\end{equation}
with mutual coherence
\begin{equation}
\Gamma_{\downarrow}(u,u') =
\gamma_\downarrow \!\left[
\varphi(u-d/2)\varphi(u'-d/2)
+\varphi(u+d/2)\varphi(u'+d/2)
\right],
\end{equation}
identical in mathematical form to the absorption kernel in Eq.~\eqref{eq:abs-kernel-2point}. For consistency with the passive fluorescence literature, we define the total scene brightness as $\varepsilon=\int\dd{u}\Gamma_{\downarrow}(u,u)=2\gamma_\downarrow$. As in the absorptive case, we can diagonalize the mutual coherence into symmetric and antisymmetric modes, with eigenvalues $\Gamma_{\downarrow,\pm}$. This formally decouples the multimode channel into two independent (quantum-limited) amplifier channels of gain $G_{\pm}\approx 1+ \Gamma_{\downarrow,\pm}$. In the subdiffraction regime ($d\ll \sigma$), the antisymmetric mode carries all information about the emitter separation, whose eigenvalue for a Gaussian PSF coincides with the HG1 coefficient, $\Gamma_{\downarrow,-}=\Gamma_{\downarrow,11}=\varepsilon d^2/(16\sigma^2)$.

Using the general QFI for a quantum-limited amplifier channel of gain $G_{-}\approx 1+\Gamma_{\downarrow,-}$~\cite{Aspachs2010:UnruhEst} (see also Appendix~\ref{app:channel-estimation}), we obtain the ultimate limit for estimating the emitter separation via fluorescence,
\begin{equation}
F_Q\bigl(d/\sigma\bigr) \approx \frac{\varepsilon(1+N_S)}{4},
\end{equation}
where $N_S$ is the total probe photon number and $\sigma$ is the PSF width. This exceeds the passive fluorescence bound~\cite{Tsang2016Superresolution} by a factor of $(1+N_S)$, showcasing the quantum enhancement to be gained through active imaging with optimal quantum probes; see Appendix~\ref{app:classical-limits} below and Table~\ref{tab:qfi_loss_gain_imaging} of the main text for a compact reference. In fact, our structured twin-beam echo protocol (as well as structured Fock-state probes) attains this bound, similar to the case of subdiffraction absorption imaging. This originates from their optimal use in amplification estimation~\cite{Aspachs2010:UnruhEst}. 

Single-mode squeezed vacuum probes are also optimal in the weak-amplification limit, but as discussed in Section~\ref{sec:single-sqz}, the use of single-mode squeezed vacuum is severely limited when both absorption and fluorescence features (or a small amount of Gaussian noise) are present.

\subsection{``Classical" limits with coherent states}
\label{app:classical-limits}

For comparison with the quantum-optimal bounds, we evaluate the achievable sensitivity limits of coherent-state probes for arbitrary quantum measurements. We focus on the canonical single-mode tasks of pure-loss and quantum-limited amplifier estimation, from which the corresponding subdiffraction imaging limits directly follow. We primarily focus on pure coherent states for simplicity but argue that the conclusions extend generally to any ``classical'' state with a positive Glauber-Sudarshan $P$-function.

Coherent states are Gaussian states, for which general expressions for the QFI are known~\cite{Pinel2013:QFIgaussian}. The QFI for a general single-mode Gaussian state, $\rho_\mathscr{G}$, with mean quadrature vector $\vec{\mu}$ and covariance $\bm{\Sigma}$ (normalized so that the vacuum has $\bm{\Sigma}_{\rm vac}=\bm I_2$; see Appendix~\ref{app:cov-bkgrd}), splits into two separate contributions,
\begin{equation}\label{eq:qfi-gauss}
    F_Q(\theta;\rho_\mathscr{G})=F_Q(\theta;\bm{\Sigma})+F_Q(\theta;\vec{\mu}).
\end{equation}
The term $F_Q(\theta;\vec{\mu})=2(\partial_\theta\vec{\mu})\bm{\Sigma}^{-1}(\partial_\theta\vec{\mu})$ contains all the information about $\theta$ that is encoded in the mean, while $F_Q(\theta;\bm{\Sigma})$ contains all the information about $\theta$ that is encoded in the covariance matrix. We do not need the general form of the covariance term $F_Q(\theta;\bm{\Sigma})$ here. Instead, we intuitively argue what explicit form it should take for loss and gain estimation with coherent-state probes. 

A coherent state has a non-zero mean, while the covariance is that of the vacuum, $\bm{\Sigma}_{\rm coh}=\bm{\Sigma}_{\rm vac}=\bm I$---hence why a coherent state is oft-called a displaced vacuum state. However, under general Gaussian channels, the output covariance may deviate from the vacuum and depend on the parameter $\theta$, so $F_Q(\theta;\bm{\Sigma})$ may contribute non-trivially to the QFI. In what follows, we describe a coherent state by the complex mean $\alpha$ (with $\vec{\mu}^{\,2}/2=|\alpha|^2$ being the average photon number) rather than the quadrature mean $\vec{\mu}$. 

\paragraph*{\textbf{Loss estimation.}} 
A coherent state with (complex) mean $\alpha$ exits a pure-loss channel with transmittance $0\leq\eta\leq 1$ as a coherent state with reduced mean $\sqrt{\eta}\alpha$ and unchanged covariance. Thus, for this special case, $F_Q^{\rm coh}(\eta;\bm{\Sigma})=0$. On the other hand, it is straightforward to show that the mean QFI term is simply $F_Q^{\rm coh}(\eta;\alpha)=|\alpha|^2/\eta$, and, therefore, $F_Q^{\rm coh}(\eta)=\frac{|\alpha|^2}{\eta}$. Reparametrizing with $\eta=e^{-\gamma_{\rm loss}}$, we find the coherent-state QFI for the absorption rate:
\begin{equation}\label{eq:qfi-loss-coherent}
    F_Q^{\rm coh}(\gamma_{\rm loss})=e^{-\gamma_{\rm loss}} |\alpha|^2,
\end{equation}
which agrees with the coherent-state row for loss estimation in Table~\ref{tab:qfi_loss_gain_imaging} to first order in $\gamma_{\rm loss}$. In the weak-loss regime ($\gamma_{\rm loss}\ll 1$), comparing Eqs.~\eqref{eq:qfi-loss} and~\eqref{eq:qfi-loss-coherent} shows that $F_Q^{\rm coh}/F_Q^{\rm opt} \sim \mathcal{O}(\gamma_{\rm loss} |\alpha|^2 / N_S)$. Consequently, coherent states require a large photon number to surpass the optimal quantum strategy, namely $|\alpha|^2 \gtrsim N_S/\gamma_{\rm loss}$.

\paragraph*{\textbf{Gain estimation.}}
For a quantum-limited amplifier of gain $G\geq 1$, a coherent-state input becomes a displaced thermal state with mean $\sqrt{G}\alpha$ and covariance $\bm{\Sigma}=(2G-1)\bm I$. To begin, consider only the covariance contribution to the QFI. Then, estimating the gain $G$ is formally equivalent to estimating the mean photon number of a thermal state with $\bar{n}_{\rm th}=G-1$ photons, with QFI $F_Q(\bar{n}_{\rm th};\bm{\Sigma})=1/[\bar{n}_{\rm th}(\bar{n}_{\rm th}+1)]$. This implies that $ F_Q(G;\bm{\Sigma})=1/[G(G-1)]$. The mean-field contribution to the QFI is straightforward to calculate: $F_Q^{\rm coh}(G;\alpha)=|\alpha|^2/[G(2G-1)]$. Adding these contributions together, it follows that
\begin{equation}
    F_Q^{\rm coh}(G)=\frac{1}{G(G-1)}+ \frac{|\alpha|^2}{G(2G-1)}.
\end{equation}
Reparametrizing $G=e^{\gamma_{\rm amp}}$, we compute the coherent-state QFI for the amplification rate $\gamma_{\rm amp}$ to be
\begin{equation}\label{eq:coh-qfi-gain}
    F_Q^{\rm coh}(\gamma_{\rm amp})=\frac{e^{\gamma_{\rm amp}}}{e^{\gamma_{\rm amp}}-1} + \frac{e^{\gamma_{\rm amp}}|\alpha|^2}{2e^{\gamma_{\rm amp}}-1},
\end{equation}
consistent with Table~\ref{tab:qfi_loss_gain_imaging} to first order in $\gamma_{\rm amp}$. In the weak-amplification limit ($\gamma_{\rm amp}\ll 1$), the vacuum contribution (i.e., the covariance QFI term) dominates, with the coherent amplitude adding only a small correction. Notably, the first term in Eq.~\eqref{eq:coh-qfi-gain} coincides with the ultimate QFI bound in Eq.~\eqref{eq:qfi-gain} for $N_S=0$---confirming that the vacuum has non-trivial metrological content for amplification estimation.

\paragraph*{\textbf{Subdiffraction imaging.}}
The extension to subdiffraction imaging with coherent-state probes follows directly through the symmetric/antisymmetric mode decompositions discussed in the preceding sections. The QFI behavior for the separation parameter $d$ inherits its traits from that of loss or gain estimation in the perturbative regime, with the feeble absorption and fluorescence rates written in the symmetric mode as, $\Gamma_{\downarrow,+}=\varepsilon\left(1- d^2/(16\sigma^2)\right)$ and $\Gamma_{\uparrow,+}=2\gamma_\uparrow \left(1- d^2/(16\sigma^2)\right)$, and antisymmetric mode as, $\Gamma_{\downarrow,-}=\varepsilon d^2/(16\sigma^2)$ and $\Gamma_{\uparrow,-}=\gamma_\uparrow d^2/(8\sigma^2)$, respectively. Coherent-state probes therefore yield finite but far-from-optimal precision relative to Fock-state or twin-beam probes. In point of fact, Eq.~\eqref{eq:qfi-loss-coherent} implies $F_Q^{\rm coh}(\Gamma_{\uparrow,\pm})\approx (1-\Gamma_{\uparrow,\pm})|\alpha|^2$. Consider populating both the symmetric and antisymmetric modes with coherent states of $\abs{\alpha}^2=N_S/2$ photons each, where $N_S$ is the total mean photon number. Then, by the chain rule and additivity, $F_Q^{\rm coh}(d/\sigma)= \sum_{k\in\{+,-\}}\sigma^2(\partial_d \Gamma_{\uparrow,k})^2 F_Q^{\rm coh}(\Gamma_{\uparrow,k}) \approx |\alpha|^2 \gamma_{\uparrow}^2 d^2/(8\sigma^2)=N_S\gamma_{\uparrow}^2 d^2/(16\sigma^2)$, as we display in Table~\ref{tab:qfi_loss_gain_imaging}. Therefore, coherent-state probes \emph{fundamentally} suffer from Rayleigh's curse in absorption imaging because, unlike fluorescence imaging, there is no intrinsic vacuum contribution that leads to a non-zero covariance QFI term. Furthermore, since any classical probe state is a convex mixture of coherent states (positive $P$ representation~\cite{Glauber1963, Sudarshan1963}), and the Fisher information is convex under mixing (cf.~\cite{Sidhu2020QFIrvw}), the coherent-state result suggests that all classical probes are Rayleigh cursed. We make this extension to arbitrary positive-$P$ classical probes rigorous below by combining a uniform multimode coherent-state bound with convexity of the QFI.

\paragraph*{\textbf{Stochastic displacement estimation.}}
For stochastic displacements $\beta \sim \mathcal{CN}(0, \gamma_{\rm agn})$, a coherent state probe is ineffective. That is, the QFI is completely independent of the coherent-state amplitude and is governed entirely by vacuum fluctuations, such that $F_Q^{\rm coh}(\gamma_{\rm agn})= 1/(\gamma_{\rm agn}(1+\gamma_{\rm agn}))\approx 1/\gamma_{\rm agn}$ for $\gamma_{\rm agn}\ll 1$.

These conclusions extend to any semi-classical probe, i.e., any state admitting a nonnegative Glauber-Sudarshan $P$ representation. To demonstrate this, we first consider the single spatial mode relevant to the subdiffraction absorption problem---the HG1 mode. An arbitrary classical state of this mode can be written as
\begin{equation}
    \rho_{\rm cl}
    = \int \dd^{2}\alpha\,P(\alpha)\dyad{\alpha},
    \qquad
    P(\alpha)\geq 0,
    \qquad
    \int \dd^{2}\alpha\,P(\alpha)=1,
\end{equation}
where $P(\alpha)$ is assumed to be independent of the unknown separation $\mathfrak{s}=d/\sigma$. The corresponding mean signal photon number is
\begin{equation}
    N_S = \int \dd^{2}\alpha\,P(\alpha)\abs{\alpha}^{2}.
\end{equation}
For two identical absorbers and the Gaussian PSF considered here, the antisymmetric mode undergoes an effective pure-loss channel with
\begin{equation}
    \Gamma_{\uparrow,-}(\mathfrak{s})
    = \gamma_{\uparrow}\left(1-e^{-\mathfrak{s}^{2}/8}\right)
    = \frac{\gamma_{\uparrow}}{8}\mathfrak{s}^{2}
    + \order{\mathfrak{s}^{4}},
\end{equation}
and transmittance $\eta_{-}(\mathfrak{s})=e^{-\Gamma_{\uparrow,-}(\mathfrak{s})}$. From Eq.~\eqref{eq:qfi-loss-coherent} and the QFI reparametrization rule, a coherent-state component therefore has
\begin{align}
    F_Q\!\left(\mathfrak{s};
    \mathcal{E}_{\mathfrak{s}}(\dyad{\alpha})\right)
    &=
    \abs{\alpha}^{2}\eta_{-}(\mathfrak{s})
    \left[
    \partial_{\mathfrak{s}}\Gamma_{\uparrow,-}(\mathfrak{s})
    \right]^{2}
    \\
    &=
    \frac{\abs{\alpha}^{2}\gamma_{\uparrow}^{2}}{16}
    \mathfrak{s}^{2}
    + \order{\mathfrak{s}^{4}}.
\end{align}
Since the mixing distribution $P(\alpha)$ is independent of $\mathfrak{s}$, linearity of the channel and convexity of the QFI imply
\begin{align}
    F_Q\!\left(
    \mathfrak{s};
    \mathcal{E}_{\mathfrak{s}}(\rho_{\rm cl})
    \right)
    &\leq
    \int \dd^{2}\alpha\,P(\alpha)
    F_Q\!\left(
    \mathfrak{s};
    \mathcal{E}_{\mathfrak{s}}(\dyad{\alpha})
    \right)
    \\
    &=
    N_S\eta_{-}(\mathfrak{s})
    \left[
    \partial_{\mathfrak{s}}\Gamma_{\uparrow,-}(\mathfrak{s})
    \right]^{2}
    \\
    &=
    \frac{N_S\gamma_{\uparrow}^{2}}{16}
    \mathfrak{s}^{2}
    + \order{\mathfrak{s}^{4}}.
    \label{eq:classical-rayleigh-bound}
\end{align}
Thus, for fixed $N_S$,
\begin{equation}
    \lim_{\mathfrak{s}\rightarrow 0}
    F_Q\!\left(
    \mathfrak{s};
    \mathcal{E}_{\mathfrak{s}}(\rho_{\rm cl})
    \right)
    =0.
\end{equation}
Because the QFI already optimizes over all measurements on the output field, no measurement can remove this vanishing sensitivity for a positive-$P$ probe. Hence, within the present two-identical-absorber model, every classical probe suffers from Rayleigh's curse in absorption imaging. Allowing $N_S$ itself to increase as $\mathfrak{s}^{-2}$ can maintain finite total information, but does not remove the vanishing information per probe photon.

For completeness, the same conclusion extends to arbitrary spatially structured classical illumination. Consider an $M$-mode positive-$P$ state
\begin{equation}
    \rho_{\rm cl}^{(M)}
    =
    \int \dd^{2M}\bm{\alpha}\,
    P(\bm{\alpha})\dyad{\bm{\alpha}},
    \qquad
    N_S
    =
    \int \dd^{2M}\bm{\alpha}\,
    P(\bm{\alpha})\norm{\bm{\alpha}}^{2}.
\end{equation}
A pure-loss channel maps a multimode coherent state to another multimode coherent state,
\begin{equation}
    \mathcal{E}_{\mathfrak{s}}(\dyad{\bm{\alpha}})
    =
    \dyad{\bm{T}_{\mathfrak{s}}\bm{\alpha}},
\end{equation}
where $\bm{T}_{\mathfrak{s}}$ is the coherent-amplitude transfer matrix. For two identical absorbers with fixed centroid, exchanging the absorbers sends $\mathfrak{s}\rightarrow-\mathfrak{s}$ while leaving the physical channel invariant. Smoothness of the Gaussian PSF therefore gives
\begin{equation}
    \bm{T}_{\mathfrak{s}}
    =
    \bm{T}_{0}
    +\order{\gamma_{\uparrow}\mathfrak{s}^{2}},
    \qquad
    \norm{
    \partial_{\mathfrak{s}}\bm{T}_{\mathfrak{s}}
    }_{\rm op}
    =
    \order{\gamma_{\uparrow}\abs{\mathfrak{s}}}.
\end{equation}
The separation dependence of the symmetric and antisymmetric eigenmodes is itself perturbative and does not modify this leading scaling. In particular, for the antisymmetric channel considered explicitly above, corrections to the eigenmode enter beyond the leading order retained in the subdiffraction expansion. Using convexity once more,
\begin{align}
    F_Q\!\left(
    \mathfrak{s};
    \mathcal{E}_{\mathfrak{s}}(\rho_{\rm cl}^{(M)})
    \right)
    &\leq
    4\int \dd^{2M}\bm{\alpha}\,
    P(\bm{\alpha})
    \norm{
    \left(
    \partial_{\mathfrak{s}}\bm{T}_{\mathfrak{s}}
    \right)\bm{\alpha}
    }^{2}
    \\
    &\leq
    4N_S
    \norm{
    \partial_{\mathfrak{s}}\bm{T}_{\mathfrak{s}}
    }_{\rm op}^{2}
    \\
    &=
    \order{
    N_S\gamma_{\uparrow}^{2}\mathfrak{s}^{2}
    }.
\end{align}
Therefore, the Rayleigh-cursed scaling is not restricted to a coherent state prepared in a particular spatial mode: any positive-$P$ classical probe has vanishing QFI as $\mathfrak{s}\rightarrow0$ at fixed mean signal photon number. The explicit prefactor can depend on the spatial distribution of the probe photons, whereas the scaling $F_Q=\order*{N_S\gamma_{\uparrow}^{2}\mathfrak{s}^{2}}$ holds under the ideal subdiffraction assumptions: two identical absorbers with known centroid, a Gaussian PSF, weak absorption, $\mathfrak{s}\ll1$, a parameter-independent probe ensemble, and fixed $N_S$.

\end{document}